\documentclass[10pt,conference,letterpaper]{IEEEtran}
\IEEEoverridecommandlockouts
\usepackage[noadjust]{cite}
\usepackage{amsmath}
\usepackage{amssymb}
\usepackage{amsthm}
\usepackage{mathtools}
\usepackage{bbm}
\usepackage{algorithm}
\usepackage[noend]{algpseudocode}
\usepackage{bm}
\usepackage{comment}
\usepackage{tabularx,booktabs}
\usepackage[capitalize]{cleveref}
\usepackage[caption=false]{subfig}
\usepackage{pgfplots}
\pgfplotsset{compat=1.15}
\usepgfplotslibrary{fillbetween}
\usetikzlibrary{patterns,arrows,plotmarks}
\usepgfplotslibrary{groupplots}
\pgfdeclarelayer{background}
\pgfsetlayers{background,main}
\usetikzlibrary{automata,positioning}
\usetikzlibrary{decorations}
\usetikzlibrary{shapes.arrows}
\usetikzlibrary{tikzmark}
\usetikzlibrary{calc}
\usetikzlibrary{decorations.markings}
\algrenewcommand\algorithmicindent{10pt}
\usepgfplotslibrary{colorbrewer}

\usepackage{tikz}
\usepackage{graphicx}

\usepackage[utf8]{inputenc}

\newcommand{\E}[1]{\mathbb{E}\left[ #1 \right]} 
\newcommand{\mc}[1]{\mathcal{#1}}   
\newcommand{\mb}[1]{\mathbf{#1}}    

\usepackage[acronym]{glossaries}

\newacronym{aoi}{AoI}{Age of Information}
\newacronym{aoii}{AoII}{Age of Incorrect Information}

\newacronym{bs}{BS}{Base Station}

\newacronym{dt}{DT}{Digital Twin}
\newacronym{dl}{DL}{Deep Learning}

\newacronym{goc}{GoC}{Goal-oriented Communication}
\newacronym{goma}{GoMA}{Goal-oriented Multiple Access}
\newacronym{mga}{MGA}{Massive Goal-oriented Access}
\newacronym{goirsa}{GO-IRSA}{Goal-oriented IRSA}

\newacronym{ibr}{IBR}{Iterated Best Response}
\newacronym{irsa}{IRSA}{Irregular Repetition Slotted ALOHA}
\newacronym{iot}{IoT}{Internet of Things}

\newacronym{jscc}{JSCC}{Joint Source Channel Coding}

\newacronym[\glslongpluralkey={Markov Decision Processes}]{mdp}{MDP}{Markov Decision Process}
\newacronym{mpi}{MPI}{Modified Policy Iteration}
\newacronym{marl}{MARL}{Multi-Agent Reinforcement Learning}
\newacronym{mega}{MEGA}{MassivE Goal-oriented Access}
\newacronym{map}{MAP}{Maximum A Posteriori}
\newacronym{mse}{MSE}{Mean Square Error}

\newacronym{ne}{NE}{Nash Equilibrium}
\newacronym{nbiot}{NB-IoT}{Narrowband IoT}

\newacronym{ofdm}{OFDM}{Orthogonal Frequency-Division Multiplexing}

\newacronym[\glslongpluralkey={Partially Observable Markov Decision Processes}]{pomdp}{POMDP}{Partially Observable Markov Decision Process}
\newacronym{pmf}{PMF}{Probability Mass Function}
\newacronym{pdf}{PDF}{Probability Density Function}
\newacronym{prb}{PRB}{Physical Resource Block}
\newacronym{ppo}{PPO}{Proximal Policy Optimization}

\newacronym{qp}{QP}{Quadratic Program}
\newacronym{qclp}{QCLP}{Quadratically Constrained Linear Programming}

\newacronym{rl}{RL}{Reinforcement Learning}
\newacronym{re}{RE}{Resource Element}

\newacronym{sic}{SIC}{Successive Interference Cancellation}

\newacronym{tdd}{TDD}{Time Division Duplex}
\newacronym{tirsa}{T-IRSA}{Threshold IRSA}

\newacronym{voi}{VoI}{Value of Information}

\definecolor{msecol}{HTML}{0011af}
\definecolor{avgcol}{HTML}{8819a0}
\definecolor{varcol}{HTML}{bf418d}
\definecolor{maxcol}{HTML}{e37076}
\definecolor{cntcol}{HTML}{f9a256}
\definecolor{mafcol}{HTML}{ffd700}

\def \fwidth{0.9\columnwidth}
\def \fheight {0.44\columnwidth}
\def \twofwidth{0.45\linewidth}
\def \twofheight {0.22\linewidth}

\def \threefwidth{0.3\linewidth}
\def \threefheight {0.2\linewidth}

\definecolor{lightgray}{HTML}{999999}
\definecolor{darkgray38}{HTML}{383838}
\definecolor{color0}{HTML}{00429D}
\definecolor{color1}{HTML}{844D99}
\definecolor{color2}{HTML}{C3608E}
\definecolor{color3}{HTML}{EF8078}
\definecolor{color4}{HTML}{FFB047}

\definecolor{blue0}{HTML}{082030}
\definecolor{blue1}{HTML}{1b465e}
\definecolor{blue2}{HTML}{317090}
\definecolor{blue3}{HTML}{489dc6}
\definecolor{blue4}{HTML}{60ccff}

\begin{document}
\title{Value-Based Massive Access through Goal-Oriented Irregular Repetition Slotted ALOHA}

\author{\IEEEauthorblockN{Pietro Talli\IEEEauthorrefmark{1}, Andrea Munari\IEEEauthorrefmark{2}, Federico Mason\IEEEauthorrefmark{1}, Federico Chiariotti\IEEEauthorrefmark{1}, and Andrea Zanella\IEEEauthorrefmark{1}}
\IEEEauthorblockA{\IEEEauthorrefmark{1}Department of Information Engineering, University of Padova, Padua, Italy}
\IEEEauthorblockA{\IEEEauthorrefmark{2}Institute of Communications and Navigation, German Aerospace Center (DLR), We{\ss}ling, Germany}
\IEEEauthorblockA{Emails: pietro.talli@phd.unipd.it, andrea.munari@dlr.de,\{federico.mason,federico.chiariotti,andrea.zanella\} @unipd.it}
\thanks{The work of F. Chiariotti and A. Zanella was supported by the Smart Networks and Services Joint Undertaking (SNS JU) under the European Union’s Horizon Europe research and innovation programme under grant agreement no. 101292933 (MAGIC-6G). The work of A. Munari was supported by the Federal Ministry of Research, Technology, and Space (BMFTR) for supporting the xG‑RIC project as part of the research program Communication Systems  “Souverän. Digital. Vernetzt.” (grant number 16KIS2429K).}}

\maketitle

\begin{abstract}
The goal-oriented communication paradigm is poised to enable novel real-time applications by easing the burden on communication networks while still delivering task-relevant information.
However, efforts so far have focused on the encoding problem, while the design of medium access schemes is still in the early stages of development, especially when connectivity is to be provided to a massive number of devices, e.g., for remote monitoring. In this respect, existing goal-oriented approaches are often centralized or based on simplified underlying mechanisms, requiring unrealistic assumptions. In this work, we present the Goal-oriented Irregular Repetition Slotted ALOHA (GO-IRSA) scheme, which combines modern random access techniques with belief-based policies. GO-IRSA does not impose significant computing loads on the sensors or require frequent feedback, and it can reduce the average and worst-case error of the estimate of a distributed Wiener process by over $30\%$ with respect to the optimal centralized solution in a network with thousands of sensors, and is robust to imperfect interference cancellation and inaccurate process knowledge.
\end{abstract}
 
\glsresetall

\section{Introduction} \label{sec:intro}

\begin{tikzpicture}[remember picture, overlay]
      \node[draw,minimum width=3in] at ([yshift=-1cm]current page.north)  {This manuscript is curently under review.};
\end{tikzpicture}

\IEEEPARstart{A}{s} the sixth generation of mobile networks (6G) foresees an ever tighter integration between the physical and digital worlds~\cite{pennanen2025intelligent}, the \gls{goc} paradigm promises to break the strict separation between applications and communication networks~\cite{gunduz2022beyond}, allowing network optimization to directly target task performance.
This groundbreaking approach has significant potential in diverse scenarios: on the one hand, distributed control systems, such as robot swarms~\cite{fang2025sensing}, which need to exchange huge volumes of data with strict latency bounds; on the other hand, massive \gls{iot} deployments need to coordinate thousands of individual low-power nodes to provide timely and accurate monitoring~\cite{kalor2025wireless}.

In typical \gls{goc} settings, the physical environment has a digital counterpart in the network, i.e., a \gls{dt}~\cite{yang2024joint}, which is used to monitor and control it.
The network orchestrator can exploit this knowledge to infer which information is more useful for the end applications,
determining the \gls{voi} of each potential communication update~\cite{feng2026goal}.
In remote monitoring, this allows one to prioritize the transmissions that more strongly improve estimates of the environment state, while in control applications, \gls{goc} strategies select the action that leads the receiver to make the most accurate decisions. 
This interaction between the communication network and the control process allows for a joint optimization that can significantly improve performance over traditional technical transmission~\cite{luo2025semantic}.

The \gls{goc} approach can greatly reduce the amount of data that must be transmitted while preserving task performance~\cite{cheng2026survey} in point-to-point communication systems, e.g., by extracting semantic content from video or 3D data and removing the need to transmit the raw information. 
On the other hand, distributed uplink applications with multiple transmitters, such as vehicular networks and massive \gls{iot} systems, introduce coordination challenges that complicate the design of goal-oriented schemes.
The resulting problem, named \gls{goma}, cannot be addressed by straightforward extensions of point-to-point architectures, as interactions between multiple agents lead to game-theoretical challenges and convergence issues~\cite{chiariotti2026theory}.

Two complementary approaches to the \gls{goma} problem have emerged in the literature, namely, \emph{pull-} and \emph{push-based} access~\cite{talli2025pragmatic}.
In a pull-based system, the \gls{bs} acts as a coordinator and decides the transmission schedule centrally, entirely avoiding interference and  packet collisions. 
The pull-based \gls{goma} problem can be modeled as a single-agent \gls{pomdp}, in which the agent placed at the \gls{bs} uses the application performance as the reward function. The \gls{rl} paradigm can then be leveraged to find near-optimal communication policies~\cite{tung2021effective}, which can also adapt to changing application and communication channel statistics. 

On the other hand, push-based \gls{goma} moves transmission decisions from the \gls{bs} to the individual nodes, which can proactively decide when to access the channel depending on local observations. The nodes know the actual value of their information, and can thus avoid transmitting low-value updates, but distributing the decision-making introduces a risk of packet collisions. The balance between the higher value of individual updates and the lower channel efficiency is tilted by an additional advantage of push-based systems: as transmission outcomes are the result of distributed decisions based on the individual knowledge of each sensor, they carry \emph{implicit information} about this knowledge~\cite{xiao2022imitation}. If a certain sensor did not transmit at a given time, the other nodes can infer that its update was not valuable enough, gaining information that can be used in subsequent decisions.

The literature on push-based \gls{goma} is still in its infancy, and even state-of-the-art schemes are based on classical random access policies, such as slotted ALOHA \cite{Abramson77:PacketBroadcasting}. Overcoming the well-known limitations of these basic schemes is often computationally complex~\cite{chiariotti2026theory} and requires very frequent feedback from the \gls{bs}~\cite{chiariotti2026goal}, limiting the potential applicability of goal-oriented enhancements based on \gls{voi}~\cite{topbas2025goal}. However, medium access research has recently developed advanced uncoordinated schemes that can mitigate these issues and improve communication performance in large networks, often grouped under the umbrella of unsourced multiple access or modern random access~\cite{Liva24_Survey}. An interesting example in this direction is \gls{irsa}, a frame-based protocol in which transmitting nodes send multiple replicas of their packets, allowing the \gls{bs} to resolve packet collisions through \gls{sic}. Although the scheme is considered a state-of-the-art random access technique for several 6G-oriented scenarios, including massive \gls{iot} systems, and its main principles have been recently embraced in 3GPP Release~19 for \gls{nbiot}~\cite{Munari26_CommStd}, its combination with \gls{goc} principles is still unexplored in the current literature.

In this work, we propose a novel push-based \gls{goma} framework that exploits \gls{irsa} to overcome the two main disadvantages of existing schemes: namely, it can limit the amount of feedback needed in a massive access scenario to a realistic value and greatly simplify the decision-making at the sensors, while increasing the performance gains over pull-based approaches. In our scheme, dubbed \gls{goirsa}, the \gls{bs} maintains a belief over the accuracy of its estimates, dynamically determining a value threshold and transmitting it to the sensor nodes, which then independently compare it to their own \gls{voi} to make transmission decisions. To the best of our knowledge, this work represents the first integration of \gls{goma} principles with massive random access, as well as one of the first push-based works to consider a timing dimension.
Our main contributions are the following:
\begin{itemize}
    \item We define \gls{goirsa}, a novel push-based \gls{goma} model that exploits \gls{irsa} as an underlying medium access mechanism, where a group of sensor nodes needs to transmit local observations about a physical environment to a central \gls{bs};
    \item We design a belief update mechanism that allows the \gls{bs} to exploit implicit information to reduce uncertainty over environment estimation;
    \item We optimize transmission decisions of the sensor nodes based on a \gls{voi}-based threshold, computed by the \gls{bs} and then broadcast to all network nodes;
    \item We analyze the performance of the proposed \gls{goirsa} scheme, revealing susbtantial gains over pull-based schemes through extensive Monte Carlo simulations, and pinpoint key trade-offs and design principles.
\end{itemize}

The rest of the paper is organized as follows:
Sec.~\ref{sec:related} discusses the relevant literature on \gls{goma} and \gls{irsa} optimization; Sec.~\ref{sec:model} presents the target scenario and the system model; Sec.~\ref{sec:results} reports our simulation results; finally, Sec.~\ref{sec:conclusions} presents our conclusions and some avenues for future work. 


\section{Related Work}
\label{sec:related}

The concept of \gls{goc} can be traced back to W.~Weaver's preface to the first edition of Shannon's work on communication theory~\cite{shannon1949mathematical}, where the importance of transmitting information is related to its relevance for the final application rather than to the faithful delivery of every bit.
An intermediate step towards this vision is represented by \emph{semantic communication}~\cite{yang2023semantic}, which focuses the optimization on the encoding: the most common example is represented by \gls{jscc} schemes~\cite{gunduz2024joint}, where the encoder and decoder are jointly optimized to preserve the semantic content of the transmitted information.
This approach has demonstrated significant performance gains across several communication tasks, but is primarily designed for point-to-point links~\cite{tung2021effective}. While multi-point coding approaches~\cite{yuan2022design, hassanpour2024multi} based on the information bottleneck principle exist, they require a joint training over all transmitters and are generally vulnerable to variations in the channel statistics.

The integration of the \gls{goc} paradigm with multiple access schemes, namely \gls{goma}, remains a relatively unexplored research direction. The most common and straightforward approach in this respect is to adopt a \emph{pull-based} architecture, in which a central \gls{bs} maintains a \gls{dt} of the entire system and assigns transmission priorities according to its estimate of the \emph{\gls{voi}} associated with each node~\cite{chen2025goal}.
Notably, most existing studies on pull-based scheduling optimize communication following proxy metrics such as \gls{aoi} or its variants, rather than application-level objectives.
In this regard, \cite{akar2024query} proposes a query-based sampling framework that models the information freshness associated with multiple remote transmitters.
More recent schemes directly target estimation~\cite{holm2023goal} and control~\cite{ayan2024optimal} performance, learning the optimal query strategy through \gls{rl} or dynamic programming~\cite{agheli2026pull}.

As discussed in Sec. \ref{sec:intro}, pull-based schemes have a fundamental limit, as they can only exploit the statistical knowledge at the \gls{bs}, represented by the \gls{dt}. On the other hand, push-based schemes are able to use the actual information available to individual sensors, ensuring a higher value for updates, but also typically incurring a significant coordination burden.
Event-triggered transmissions are considered in~\cite{wang2021value}, analyzing the effect of basic on-off mechanisms corresponding to value thresholds. A similar model is analyzed in~\cite{wu2024goal}, which also includes retransmissions. However, neither work explicitly addresses contention over resource-limited channels.

A comprehensive push-based \gls{goma} framework is proposed in~\cite{agheli2024goal}, where multiple sensing agents observe a common process and optimize transmission decisions according to the expected effectiveness of the information.
This line of research is further extended in~\cite{agheli2025integrated}, which combines push-based sensing updates with pull-based query mechanisms.
Another recent work~\cite{chiariotti2026theory} generalizes the framework considering arbitrary \gls{voi} belief distributions, moving towards a more complete theory, but the scalability of the approach is limited by the need for the strategy to be computed by each node at every time step.
A simpler belief-based approach was developed in~\cite{chiariotti2026goal}, which examines belief updates over slotted ALOHA, obtaining good results in wider networks. However, the analysis in both papers neglects the impact and overhead cost of the downlink feedback channel, which is required at every time step.

Recent years have also seen the first steps towards leveraging modern random access schemes for \gls{goc}. Initial works evaluated the performance of \gls{irsa} in terms of the \gls{aoi} \cite{munari2021modern}, possibly in combination with an age-based threshold~\cite{asgari22}, identifying a fundamental trade-off driven by the frame duration, with longer contention intervals allowing for higher throughput while possibly stalling delivery of newly generated packets. Similar effects were observed for other modern random access schemes, such as frameless ALOHA \cite{Munari23_TCOM}, and recent contributions have also started considering asynchronous (unslotted) replica transmissions~\cite{Uysal26_ASTRA}. 

In turn, fewer works explored the potential of \gls{irsa} towards \gls{goc} beyond \gls{aoi}. An interesting example is provided by \cite{mubarak2026on}, where a \gls{tirsa} scheme based on the \gls{aoii} was proposed, providing some approximated analysis of the system performance. Notably, the approach does not rely on any distributed knowledge or belief calculation, but rather relies on a fixed and pre-calculated threshold. Another recent work~\cite{zhang2024value} considers an extension of \gls{irsa} in which multiple \gls{iot} nodes are assigned different priority classes, setting higher transmission probabilities for more urgent users. Although this approach progresses towards the integration of \gls{goc} into the medium access layer, the proposed metric is not directly related to the contribution of each packet to the final application objective. Hence, to our knowledge, the extension of \gls{irsa}-based methodologies to the \gls{goma} problem remains an open research direction.

\section{System Model}
\label{sec:model}

We consider a scenario in which a set $\mathcal N$ of $N$ sensor nodes report data to a \gls{bs} over a shared wireless channel. Each node $n\in\mathcal N$ measures the state of an independent Wiener process $x_n$~\cite{wiener1921average}, which the \gls{bs} aims to track through the use of a \gls{dt}.
Consequently, if we know the value of the process at time $t$, its value at time $t+\tau$ is normally distributed:
\begin{equation}\label{eq:update} x_n(t+\tau)\sim\mc{N}\left(x_n(t),\tau\sigma_n^2\right),
\end{equation}
where $\sigma_{n}$ is a scaling parameter.
Time is discretized into slots of duration $T_s$, which are organized into \emph{frames} of equal duration $T_f$, each covering an integer number of slots $S=T_f/T_s$. We use index $j\in\{1,\ldots,S\}$ for slots and $k\in\mathbb{N}$ for frames, so that the $j$-th slot of frame $k$ begins at time $(j-1)T_s+kT_f$. The state of process $x_n$ at the beginning of slot $j$ in frame $k$, which is observed directly and noiselessly by sensor $n$, is denoted as $x_n(j,k)$.

\subsection{Application Model}

We consider the \gls{dt} for each process to be a belief distribution, updated at each slot and identified by \gls{pdf} $\varphi_{n;j,k}\in\Omega(\mathbb{R})$, where $\Omega(\cdot)$ denotes the probability simplex over a set. We denote the \gls{map} estimate resulting from the belief distribution $\varphi_{n;j,k}$ by $\hat{x}_n(j,k)$ and define the \emph{drift error} $e_n(j,k)$ as the squared difference between the real state of the system and the \gls{map} estimate:
\begin{equation}
\label{eq:drift}
    e_n(j,k)=|x_n(j,k)-\hat{x}_n(j,k)|^2.
\end{equation}
We also note that, as the \gls{bs} receives observations of a Wiener process, its belief distribution is symmetrical, and the \gls{map} estimator corresponds to its mode. The error of the system $E(k)$ is measured through the \gls{mse}, i.e.,
\begin{equation}
\label{eq:mse}
    E(k)=\frac{1}{SN}\sum_{j=1}^S\sum_{n=1}^N e_n(j,k).
\end{equation}

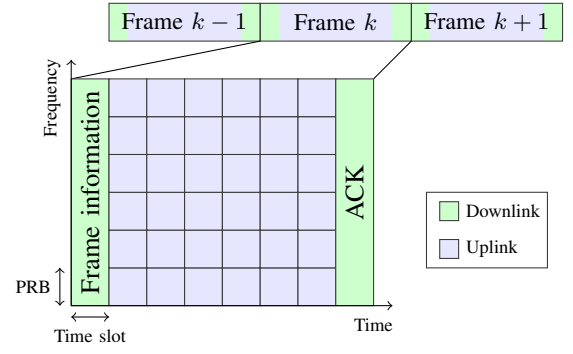
\begin{figure}[t]
    \centering
    \begin{tikzpicture}

\node[rectangle, fill=green!20,at={(0.375cm,3.5cm)}, minimum width=0.25cm,minimum height=0.5cm]{};
\node[rectangle, fill=blue!10,at={(1.25cm,3.5cm)}, minimum width=1.5cm,minimum height=0.5cm]{};
\node[rectangle, fill=green!20,at={(2.25cm,3.5cm)}, minimum width=0.5cm,minimum height=0.5cm]{};
\node[rectangle, fill=blue!10,at={(3.25cm,3.5cm)}, minimum width=1.5cm,minimum height=0.5cm]{};
\node[rectangle, fill=green!20,at={(4.25cm,3.5cm)}, minimum width=0.5cm,minimum height=0.5cm]{};
\node[rectangle, fill=blue!10,at={(5.25cm,3.5cm)}, minimum width=1.5cm,minimum height=0.5cm]{};
\node[rectangle, fill=green!20,at={(6.125cm,3.5cm)}, minimum width=0.25cm,minimum height=0.5cm]{};

\node[rectangle, draw=darkgray38,at={(1.25cm,3.5cm)}, minimum width=2cm,minimum height=0.5cm]{};
\node[rectangle, draw=darkgray38,at={(3.25cm,3.5cm)}, minimum width=2cm,minimum height=0.5cm]{};
\node[rectangle, draw=darkgray38,at={(5.25cm,3.5cm)}, minimum width=2cm,minimum height=0.5cm]{};
\node[at={(1.25cm,3.5cm)}]{\small Frame $k-1$};
\node[at={(3.25cm,3.5cm)}]{\small Frame $k$};
\node[at={(5.25cm,3.5cm)}]{\small Frame $k+1$};

\draw (2.25cm,3.25cm) -- (-0.25cm,2.75cm);
\draw (4.25cm,3.25cm) -- (3.75cm,2.75cm);

\node[rectangle, fill=green!20,draw=darkgray38,at={(0cm,1.25cm)}, minimum width=0.5cm,minimum height=3cm]{};
    
    \foreach \i in {1,...,6} {
        \foreach \j in {0,...,5} {
            \node[rectangle, fill=blue!10,draw=darkgray38,at={(0.5*\i cm,0.5*\j cm)}, minimum width=0.5cm,minimum height=0.5cm]{};
        }
    }

\node[rectangle, fill=green!20,draw=darkgray38,at={(3.5cm,1.25cm)}, minimum width=0.5cm,minimum height=3cm]{};

\node[at={(3.5,1.25)},rotate=90]{ACK};
\node[at={(0,1.25)},rotate=90]{Frame information};

\draw[<->] (-0.4,-0.25) -- (-0.4,0.25);
\node[at={(-0.75,0)}] {\scriptsize PRB};
\draw[<->] (-0.25,-0.4) -- (0.25,-0.4);
\node[at={(0,-0.65)}] {\scriptsize Time slot};

\draw[->] (-0.25,-0.25) -- (-0.25,3);
\draw[->] (-0.25,-0.25) -- (4,-0.25);
\node[at={(3.75,-0.5)}] {\scriptsize Time};
\node[at={(-0.5,2.5)},rotate=90] {\scriptsize Frequency};

\node[rectangle,draw=darkgray38,at={(5.25,0.75)}, minimum width=1.6cm,minimum height=1cm]{};
\node[rectangle, fill=blue!10,draw=darkgray38,at={(4.75,0.5)}, minimum width=0.25cm,minimum height=0.25cm]{};
\node[rectangle, fill=green!20,draw=darkgray38,at={(4.75,1)}, minimum width=0.25cm,minimum height=0.25cm]{};
\node[at={(4.85,1)},anchor=west] {\scriptsize Downlink};
\node[at={(4.85,0.5)},anchor=west] {\scriptsize Uplink};

\end{tikzpicture}
    \caption{Frame structure.}
    \label{fig:frame_grid}
\end{figure}

\subsection{Communication Model}

We model the communication system using a common frame structure for push-pull access~\cite{pandey2026pullpushmag}, shown in Fig.~\ref{fig:frame_grid}: each frame has a fixed duration, and follows the \gls{ofdm} structure, with $S$ time slots and $F$ subcarriers. The duration of a time slot for the application model coincides with the communication time slot. We adopt a \gls{tdd} approach, in which all but the first and last symbol of each frame are reserved for uplink communication: the total number of \glspl{re} available to the sensors is thus $M=(S-2)F$, and each uplink packet transmission can fit into a single \gls{re}.

At the end of each frame, the \gls{bs} broadcasts a collective acknowledgment packet to all nodes: this is a simple binary vector with $N$ entries, corresponding to a decoded/not-decoded outcome for each node. On the other hand, at the beginning of the frame, the \gls{bs} sends a downlink packet, which contains control information for the uplink part. We thus distinguish between pull- and push-based access: in the former, the frame information packet specifies the schedule, i.e., exactly which node should transmit over which \gls{re}, and thus requires a payload of $
M\log_2(N)$ bits. The sensors then behave as the packet specifies, and access is fully orthogonal. On the other hand, in push-based access, the \gls{bs} may specify a probability or condition to transmit, and sensors will independently decide whether to do so and over which \glspl{re}. In the remainder, we assume a collision channel model \cite{Abramson77:PacketBroadcasting}: whenever two or more packets are present over one slot, none of them can be decoded, whereas a slot containing a single packet (singleton) always leads to retrieval of the information content. The model has been widely adopted in the literature both for \gls{goma}~\cite{agheli2025integrated,chiariotti2026theory} and \gls{irsa}~\cite{mubarak2026on,Paolini15:TIT_CSA} in view of its tractability, and captures the fundamental trade-offs of medium contention.

\subsection{Irregular Repetition Slotted ALOHA (IRSA)}

When considering push-based schemes, we will focus in particular on~\gls{irsa}~\cite{DeGaudenzi07:CRDSA,Liva11:IRSA,Paolini15:TIT_CSA}.
In this scheme, a node which decides to send data over a frame transmits multiple identical copies (replicas) of its message over some of the available REs. The number of replicas is drawn independently by each transmitter from a probability distribution, defined in advance and available to all network nodes. Following a common notation, the distribution is indicated as $\Lambda(x) = \sum_{\ell=1}^L \Lambda_\ell x^\ell$, where $\sum_\ell \Lambda_\ell = 1$, $\Lambda_\ell$ denotes the probability for the device to send exactly $\ell$ replicas, and $L$ is the maximum number of copies that can be transmitted. To distribute its $\ell$ replicas across the available \glspl{re}, a device randomly selects $\ell$ of the $S-2$ uplink slots in the frame, choosing one of the $F$ subcarriers for each selected slot. We group the subcarrier and time indexes chosen by node $n$ at frame $k$ in vectors $\mb{f}_n(k)$ and $\mb{s}_n(k)$, with components in $\{1,\ldots,F\}$ and $\{2,\ldots,S-1\}$, respectively.

Every replica further contains a pointer to the REs where all other copies of the same message are sent, e.g., by means of a short header, or with more advanced solutions~\cite{Munari26_CommStd}. At the receiving end, decoding is performed at the end of the frame, relying on \gls{sic}. Specifically, the receiver begins by identifying slots containing a single packet (singleton) and decoding them. Once a packet is decoded, its waveform is reconstructed and subtracted from the overall incoming signal across the other \glspl{re} where it was transmitted, eliminating its interference contribution. This cancellation may uncover new singleton slots, enabling the decoding of packets that previously suffered collisions. This procedure is iterated until all messages are successfully retrieved or no further decoding is possible. In the latter scenario, some transmitting nodes remain unresolved; in this case, we assume the \gls{bs} can estimate neither their number nor their identities.

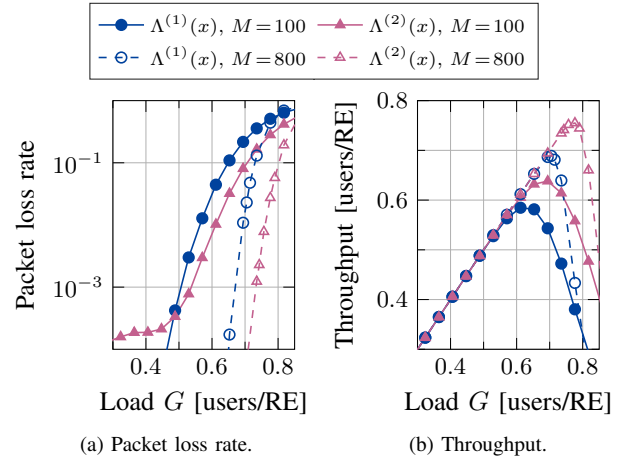
\begin{figure}
    \centering
    \centering
    \subfloat{
\begin{tikzpicture}

\definecolor{darkgray176}{RGB}{176,176,176}

\begin{axis}[
width=0cm,
height=0.1cm,
scale only axis,
tick align=inside,
xmin=0, xmax=0,
ymin=0, ymax=0.1,
legend style={
    draw=white!15!black,
    font=\scriptsize, at={(0, 0)}, anchor=south,
    /tikz/every even column/.append style={column sep=0.2em}
},
legend columns=2
]

\addplot[color0,semithick,mark=*] table {
0    0.1
};

\addplot[color2, semithick,mark=triangle*] table {
0    0.1
};
\addplot[color0,dashed,semithick,mark=o,mark options={solid}] table {
0    0.1
};

\addplot[color2,dashed, semithick,mark=triangle,mark options={solid}] table {
0    0.1
};

\legend{$\Lambda^{(1)}(x){,}\ M\!=\!100$,$\Lambda^{(2)}(x){,}\ M\!=\!100$,
$\Lambda^{(1)}(x){,}\ M\!=\!800$,$\Lambda^{(2)}(x){,}\ M\!=\!800$
}
\end{axis}
\end{tikzpicture}} \\ 
    \vspace{-0.3cm}    \setcounter{subfigure}{0}
    \subfloat[Packet loss rate.\label{fig:plr_irsa}]{
\begin{tikzpicture}

\definecolor{darkgray176}{RGB}{176,176,176}

\begin{axis}[
width=\twofwidth,
height=0.55\linewidth,
tick label style={font=\small},
grid=both,
grid style={line width=.1pt, draw=gray!10},
major grid style={line width=.2pt,draw=gray!40},
minor tick num=1,
xlabel={Load $G$ [users/RE]},
ylabel={Packet loss rate},
x grid style={darkgray176},
xmin=0.3, xmax=0.85,
xtick style={color=black},
y grid style={darkgray176},
ymin=1e-4, ymax=1,
ytick style={color=black},
ymode=log,
legend style={draw=white!15!black, font=\scriptsize, at={(0.8, 0.98)}, anchor=north east, /tikz/every even column/.append style={column sep=0.2em}},
legend columns=2
]

\addplot[color=color0,semithick,mark=*] table[x index=0,y index=2] {plot/plr_12_slot.txt};

\addplot[color=color2,semithick,mark=triangle*] table[x index=0,y index=1] {plot/plr_12_slot.txt};

\addplot[color=color0,semithick,mark=o,dashed,mark options={solid}] table[x index=0,y index=1] {plot/plr_82_slot_regular.txt};

\addplot[color=color2,semithick,mark=triangle,dashed,mark options={solid}] table[x index=0,y index=1] {plot/plr_82_slot_irregular.txt};

\end{axis}
\end{tikzpicture}}
    \subfloat[Throughput.\label{fig:tru_irsa}]{
\begin{tikzpicture}

\definecolor{darkgray176}{RGB}{176,176,176}

\begin{axis}[
width=\twofwidth,
height=0.55\linewidth,
tick label style={font=\small},
grid=both,
grid style={line width=.1pt, draw=gray!10},
major grid style={line width=.2pt,draw=gray!40},
minor tick num=1,
xlabel={Load $G$ [users/RE]},
ylabel={Throughput [users/RE]},
x grid style={darkgray176},
xmin=0.3, xmax=0.85,
xtick style={color=black},
y grid style={darkgray176},
ymin=0.3, ymax=0.8,
ytick style={color=black},
legend style={draw=white!15!black, font=\scriptsize, at={(0.8, 0.98)}, anchor=north east, /tikz/every even column/.append style={column sep=0.2em}},
legend columns=2
]

\addplot[color=color0,semithick,mark=*] table[x index=0,y index=2] {plot/tru_12_slot.txt};

\addplot[color=color2,semithick,mark=triangle*] table[x index=0,y index=1] {plot/tru_12_slot.txt};

\addplot[color=color0,semithick,mark=o,dashed,mark options={solid}] table[x index=0,y index=1] {plot/tru_82_slot_regular.txt};

\addplot[color=color2,semithick,mark=triangle,dashed,mark options={solid}] table[x index=0,y index=1] {plot/tru_82_slot_irregular.txt};

\end{axis}
\end{tikzpicture}}
    \caption{\gls{irsa} performance using distributions $\Lambda^{(1)}(x) = x^4$ and $\Lambda^{(2)}(x) = 0.86x^3 + 0.14 x^8$. Two frame lengths ($M\!=\!100$ and $M\!=\!800$) are considered, with $F\!=\!10$ in both cases. Results were obtained for Poisson traffic of parameter $G$ users/RE over each frame. Throughput is also expressed in terms of served users per \gls{re}.}
    \label{fig:basic_perf_irsa}
\end{figure}

The performance of \gls{irsa} is fundamentally driven by the degree distribution $\Lambda(x)$ and by the number $M$ of \glspl{re} available for transmission. These well-known aspects are exemplified in Fig.~\ref{fig:basic_perf_irsa}, where we report the packet loss rate, i.e., the probability for a transmitting user not to be decoded after \gls{sic}, as well as the throughput of the scheme against the channel load $G$. This is defined as the average number of users transmitting over a frame divided by the number of available \glspl{re} $M$, and is thus independent of the selected distribution. The results were generated considering two replica distributions: $\Lambda^{(1)}(x)=x^4$, i.e., all nodes transmit $4$ replicas; and $\Lambda^{(2)}(x)=0.86x^3 + 0.14x^8$, originally proposed in~\cite{ivanov2017broadcast} as a good solution to operate with small to average frame sizes. In the plots, $F=10$ subcarriers are used, comparing the performance of the scheme when $S=12$ and $S=82$, i.e., $M=100$ and $M=800$, respectively. Notably, longer frames lead to substantial improvements in terms of reliability and throughput, decreasing the probability to encounter stopping sets that halt the SIC procedures \cite{Liva11:IRSA,ivanov2017broadcast}. Operating with a larger $S$ entails a cost in terms of more sporadic transmission decisions, driving the design of effective \gls{goc} schemes, as will be discussed in Sec. \ref{sec:results}.


\section{The GO-IRSA Scheme}

In this section, we present the \gls{goirsa} solution. The basic design is simple, and requires almost no computational effort from the distributed sensors. At the beginning of frame $k$, the \gls{bs} computes a threshold $\theta(k)$, then distributes it to the sensors using the control information packet. Upon receiving this, each sensor observes the value of the monitored process, and computes its \gls{voi} $e_n(0,k)$, deciding to participate in the frame if $e_n(0,k)>\theta(k)$. If sensor $n$ decides to send an update, it generates a payload containing the reading  obtained in slot $\inf(\mb{s}_n(k))$, i.e., just before its first transmission in the frame, and then sends its replicas containing such value following the \gls{irsa} scheme. 

\subsection{Digital Twin Belief Update}

At the end of frame $k$, the \gls{bs} performs \gls{sic} over all uplink \glspl{re}, attempting to retrieve transmitted packets. As discussed, there are two possible outcomes: either all packets are successfully retrieved, or there are some unresolved collisions and part of the information is lost.

We first consider nodes that successfully transmitted in frame $k$, whose identity and transmission vectors are recovered from the packets. The \gls{bs} thus knows the state of the process $x_n(\inf(\mb{s}_n(k),k)$. However, the process keeps evolving, so that, at the beginning of the next frame, the belief \gls{pdf} $\varphi_{n;0,k+1}$ is
\begin{equation}
    \varphi_{n;0,k+1}(z)=\frac{\exp\left(-z^2(2(S-\inf(\mb{s}_n(k))T_s)^{-1}\sigma_n^{-2}\right)}{\sqrt{2\pi(S-\inf(\mb{s}_n(k))T_s\sigma_n^2}}.
\end{equation}
This corresponds to a Gaussian distribution with mean $0$ and variance $[S-\inf(\mb{s}_n(k)]T_s\sigma_n^2$, following~\eqref{eq:update}.
After receiving the acknowledgment packet, the sensor also knows that the \gls{bs} updated $\hat{x}_n(\inf(\mb{s}_n(k),k)$, and that the estimate will remain stable, as the Wiener process is a martingale.\footnote{Our formulation can be easily extended to any linear dynamical system, as long as the update equation is known to both the sensor and \gls{bs}.} It will then use the estimated value to compute the error $e_n$ in future frames.

We now consider the \emph{implicit information} in the case in which there are no unresolved collisions: as we know the identity of all nodes that transmitted in frame $k$, we also know that other nodes did not transmit. Since the decision rule that the nodes use is a threshold policy, the \gls{bs} can infer that node $m$, which did not send an update, must have had $e_m(0;k)<\theta(k)$. The set of values $\mc{X}_m(\hat{x},\theta)$ for which node $m$ chose not to transmit in frame $k$ is 
\begin{equation}
    \mc{X}_m(\hat{x},\theta)=\left[\hat{x}-\sqrt{\theta},\hat{x}+\sqrt{\theta}\right].
\end{equation}
We can then obtain an \emph{a posteriori} belief $\tilde{\varphi}_{m;0,k}$ by applying Bayes' theorem, truncating the distribution:
\begin{equation}\label{eq:posterior}
    \tilde{\varphi}_{m;0,k}=\begin{cases}
        \frac{\varphi_{m;0,k}(z)}{\zeta_m(k)}, &\text{if }z\in\mc{X}_m(\hat{x}_m(0,k),\theta(k));\\
        0, &\text{otherwise,}
    \end{cases}
\end{equation}
where value $\zeta_m(k)$ corresponds to the \emph{a priori} probability of node $m$ not transmitting:
\begin{equation}
    \zeta_m(k)=\int_{\mc{X}_m(\hat{x}_m(0,k),\theta(k))}\varphi_{m;0,k}(z) dz.
\end{equation}
We then note that the value of the process $x_m$ evolved for a full frame,so that we have
\begin{equation}\label{eq:evolution}
    x_m(0,k+1)\sim\mc{N}\left(x_m(0,k),T_f\sigma_m^2\right).
\end{equation}
In order to obtain $\varphi_m(0,k+1)$, we need to take the convolution of~\eqref{eq:posterior} and the Gaussian \gls{pdf} from~\eqref{eq:evolution}:
\begin{equation}
 \varphi_{m;0,k+1}(z)=\quad \int\displaylimits_{\mathclap{\mc{X}_m(\hat{x}_m(0,k),\theta(k))}} \quad \frac{\tilde{\varphi}_{m;0,k}(w)\exp\left(-\frac{(z-w)^2}{2T_f\sigma_m^2}\right)}{\sqrt{2\pi T_f\sigma_m^2}}dw.
\end{equation}

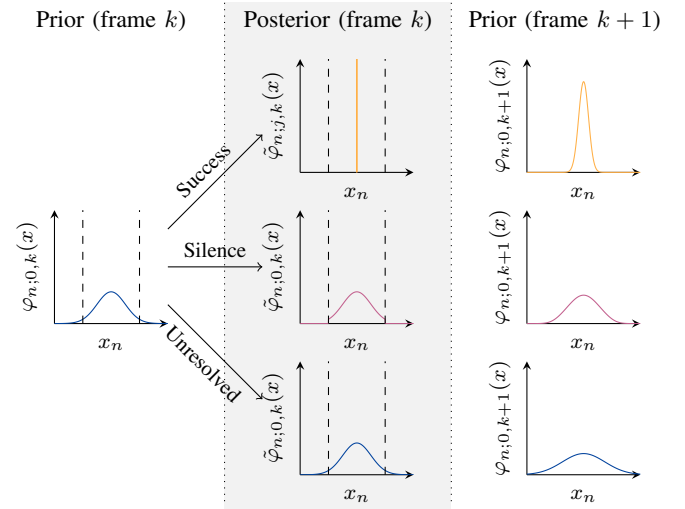
\begin{figure}
    \centering
    \begin{tikzpicture}

\node[fill=gray!10,minimum width=3cm,minimum height=6.7cm,at={(2.5cm,0.9cm)}]{};

\draw[dotted] (1cm,-2.35cm) -- (1cm,4.25cm);
\draw[dotted] (4cm,-2.35cm) -- (4cm,4.25cm);
\node[at={(-0.5cm,4cm)}]{\small Prior (frame $k$)};
\node[at={(2.5cm,4cm)}]{\small {Posterior} (frame $k$)};
\node[at={(5.5cm,4cm)}]{\small Prior (frame $k+1$)};

\begin{axis}
    [at={(-1.25cm,0)},
    width=1.5cm,
    height=1.5cm,
    scale only axis,
    name=before,
    xtick=\empty,
    ytick=\empty,
    xmin=-4 , xmax=4,
    ymin=0, ymax=1,
    label style={font=\footnotesize},     
    axis lines=left,
    xlabel={$x_n$},
    ylabel={$\varphi_{n;0,k}(x)$}
    ]
\addplot[
color=color0
]
table[x=x,y=phi_orig] {fig/belief_data.dat};
\addplot[
color=black,dashed
]
table {
-2  0
-2  1
};
\addplot[
color=black,dashed
]
table {
2  0
2  1
};

\end{axis}

\begin{axis}
    [at={(2cm,-2cm)},
    width=1.5cm,
    height=1.5cm,
    scale only axis,
    name=before,
    xtick=\empty,
    ytick=\empty,
    xmin=-4 , xmax=4,
    ymin=0, ymax=1,
         label style={font=\footnotesize},     axis lines=left,
    xlabel={$x_n$},
    ylabel={$\tilde{\varphi}_{n;0,k}(x)$}
    ]
    
\addplot[
color=color0
]
table[x=x,y=phi_orig] {fig/belief_data.dat};
\addplot[
color=black,dashed
]
table {
-2  0
-2  1
};
\addplot[
color=black,dashed
]
table {
2  0
2  1
};
\end{axis}

\begin{axis}
    [at={(5cm,-2cm)},
    width=1.5cm,
    height=1.5cm,
    scale only axis,
    name=before,
    xtick=\empty,
    ytick=\empty,
    xmin=-4 , xmax=4,
    ymin=0, ymax=1,
    label style={font=\footnotesize},
    axis lines=left,
    xlabel={$x_n$},
    ylabel={$\varphi_{n;0,k+1}(x)$}
    ]
\addplot[
color=color0
]
table[x=x,y=phi_next] {fig/belief_data.dat};

\end{axis}

\draw[->] (0.25cm,1.25cm) -- node[midway,above,rotate=45]{\footnotesize Success} (1.5cm,2.5cm);
\draw[->] (0.25cm,0.75cm) -- node[midway,above]{\footnotesize Silence} (1.5cm,0.75cm);
\draw[->] (0.25cm,0.25cm) -- node[midway,below,rotate=-45]{\footnotesize Unresolved} (1.5cm,-1cm);

\begin{axis}
    [at={(2cm,2cm)},
    width=1.5cm,
    height=1.5cm,
    scale only axis,
    name=before,
    xtick=\empty,
    ytick=\empty,
    xmin=-4 , xmax=4,
    ymin=0, ymax=1,
    label style={font=\footnotesize},     
    axis lines=left,
    xlabel={$x_n$},
    ylabel={$\tilde{\varphi}_{n;j,k}(x)$}
    ]
\addplot[
color=color4
]
table{
-0.01	0
0	1
0.01	0
};
\addplot[
color=black,dashed
]
table {
-2  0
-2  1
};
\addplot[
color=black,dashed
]
table {
2  0
2  1
};

\end{axis}

\begin{axis}
    [at={(5cm,2cm)},
    width=1.5cm,
    height=1.5cm,
    scale only axis,
    name=before,
    xtick=\empty,
    ytick=\empty,
    xmin=-4 , xmax=4,
    ymin=0, ymax=1,
    label style={font=\footnotesize},
    axis lines=left,
    xlabel={$x_n$},
    ylabel={$\varphi_{n;0,k+1}(x)$}
    ]
\addplot[
color=color4
]
table[x=x,y=phi_succ_next] {fig/belief_data.dat};
\end{axis}

\begin{axis}
    [at={(2cm,0)},
    width=1.5cm,
    height=1.5cm,
    scale only axis,
    name=before,
    xtick=\empty,
    ytick=\empty,
    xmin=-4 , xmax=4,
    ymin=0, ymax=1,
    label style={font=\footnotesize},
    axis lines=left,
    xlabel={$x_n$},
    ylabel={$\tilde{\varphi}_{n;0,k}(x)$}
    ]
\addplot[
color=color2
]
table[x=x,y=phi_trunc] {fig/belief_data.dat};
\addplot[
color=black,dashed
]
table {
-2  0
-2  1
};
\addplot[
color=black,dashed
]
table {
2  0
2  1
};
\end{axis}

\begin{axis}
    [at={(5cm,0)},
    width=1.5cm,
    height=1.5cm,
    scale only axis,
    name=before,
    xtick=\empty,
    ytick=\empty,
    xmin=-4 , xmax=4,
    ymin=0, ymax=1,
    label style={font=\footnotesize},
    axis lines=left,
    xlabel={$x_n$},
    ylabel={$\varphi_{n;0,k+1}(x)$}
    ]
\addplot[
color=color2
]
table[x=x,y=phi_trunc_next] {fig/belief_data.dat};

\end{axis}

\end{tikzpicture}\vspace{-0.4cm}
    \caption{Diagram showing the outcome of the belief update process for different outcomes of the frame for a given node $n$.}
    \label{fig:belief_update}
\end{figure}

On the other hand, if the frame is unresolved, the \gls{bs} does not know the identity or even the number of unresolved transmissions.\footnote{Estimating the number of transmissions in an unresolved frame is a complex task, as it requires enumerating possible stopping sets for \gls{sic}~\cite{moroglu2020short}, and we consider it to be beyond the scope of our work.} In this case, we do not use implicit information, and simply update the belief distribution for any node that was \emph{not} successfully decoded over frame $k$, i.e., either remained silent or could not be retrieved due to collisions, as
\begin{equation}
 \varphi_{m;0,k+1}(z)=\int_{-\infty}^{\infty} \frac{\varphi_{m;0,k}(w)\exp\left(-\frac{(z-w)^2}{2T_f\sigma_m^2}\right)}{\sqrt{2\pi T_f\sigma_m^2}}dw.
\end{equation}

Using these equations, we can update the belief distribution for all sensors at the end of each frame, refreshing the \gls{dt}. An illustrative example of the belief update process is shown in the diagram in Fig.~\ref{fig:belief_update}: if node $n$ is successful, the posterior belief for the measurement time becomes deterministic, and the prior used for the next frame is a Gaussian variable with a low variance, accounting for changes in the process since the received update. On the other hand, if the node is silent and there are no unresolved collisions in the frame, the \gls{dt} truncates the belief at the threshold (represented with dashed lines), then adds a Gaussian term to account for changes during the frame. Finally, if the frame is unresolved, the posterior belief is identical to the prior, and the prior for the next frame includes another Gaussian component, as in the silent case.

\subsection{Threshold Selection}

The other component of \gls{goirsa} is the selection of the threshold $\theta(k)$. This choice presents a trade-off: setting a high threshold reduces transmissions and increases the overall \gls{mse}, yet aiming for a low threshold increases the risk of unresolved collisions and possibly  degrades the overall performance. We thus consider vector $\bm{\varphi}_{0,k}$, including all belief distributions available at the BS at the start of  frame $k$. The threshold leading to an expected load $\bar{G}$ over frame $k$ is thus the solution of the integral equation
\begin{equation}
\label{eq:threshold}
\sum_{n=1}^N \left[1-\int_{-\sqrt{\theta}}^{\sqrt{\theta}}\varphi_{n;0,k}(z)dz\right]=\bar{G}M.
\end{equation}
We can thus simply identify a target load $\bar{G}$ and use it as a rule to determine the threshold in each step. As we will see in Sec.~\ref{sec:results}, this simple strategy is quite effective at reducing the \gls{mse} of the distributed monitoring.

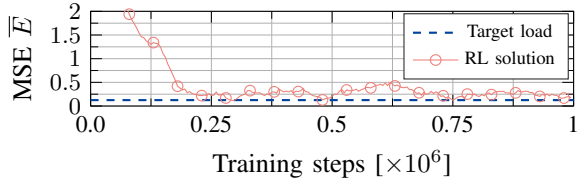
\begin{figure}
    \centering
\begin{tikzpicture}

\definecolor{darkgray176}{RGB}{176,176,176}

\begin{axis}[
width=\fwidth,
height=0.32\columnwidth,
grid=both,
tick label style={font=\small},
grid style={line width=.1pt, draw=gray!10},
major grid style={line width=.2pt,draw=gray!40},
minor tick num=1,
xlabel={Training steps [$\times 10^6$]},
ylabel={MSE $\overline{E}$},
x grid style={darkgray176},
xmin=0, xmax=200,
xtick style={color=black},
y grid style={darkgray176},
ymin=0.0, ymax=2.0,
xtick={0,50,100,150,200},
xticklabels={0.0,0.25,0.5,0.75,1},
ytick style={color=black},
legend style={
    draw=white!15!black,
    font=\scriptsize,
    at={(0.99,0.98)},
    anchor=north east,
    /tikz/every even column/.append style={column sep=0.2em}
}
]

\addplot[dashed, color0, thick] coordinates {
    (0, 0.1247784324362874)
    (200, 0.1247784324362874)
};

\addplot[color3,mark=o, mark repeat=10] table[
    x expr=\coordindex,
    y expr=-\thisrow{Value},
    col sep=comma
] {plot/ppo_training_progress.csv};

\legend{Target load, RL solution};

\end{axis}
\end{tikzpicture}
    \caption{Performance of the \gls{rl} agent during the training phase, with $M\!=\!200$.}
    \label{fig:rl_plot}
\end{figure}

In order to validate this approach, we consider an \gls{rl} solution as part of our preliminary analysis.
In this case, we place the agent at the \gls{bs}, with a state space corresponding to the possible belief distributions for all nodes, quantized as described in Sec.~\ref{sec:results}.
The action of the agent is to set a set of thresholds, one per each possible number of replicas in the \gls{irsa} distribution, with an increase in replicas corresponding to lower thresholds. The set of values is distributed in the downlink frame information, and a sensor with error larger than a given threshold contends sending the corresponding number of replicas. We report the \gls{mse} performance of the \gls{rl} agent during its training with \gls{ppo} in Fig.~\ref{fig:rl_plot}, comparing it with a system with a fixed target load $\bar{G}=0.7$, picked to minimize the \gls{mse}. The figure clearly shows that the \gls{rl} agent never outperforms the target load approach, even after convergence. In this respect, it is also interesting to notice that the use of different thresholds, aiming to favor the decoding of nodes with higher errors by having them transmit more replicas, does not play a significant role. Based on these results, the simple target load approach in \eqref{eq:threshold}  reduces complexity and training costs and does not degrade performance, and we will consider it as the main solution in the following.

\section{Simulation Results}
\label{sec:results}

In this section, we analyze the proposed \gls{goirsa} scheme by Monte Carlo simulation. We run $N_{\text{sim}}=50$ episodes of $L=200$ frames for each configuration.\footnote{The simulation code is available at https://github.com/pietro-talli/GO-IRSA} We consider a scenario with $N=2000$ sensing nodes communicating with the \gls{bs} over an \gls{ofdm} grid with $F=10$ subcarriers and time slots lasting $T_s=10$~ms. We vary the number of slots in the frame, with $S\in\{12,22,42,62,82\}$, obtaining $M\in\{100,200,400,600,800\}$ \glspl{re} for uplink communication when considering the downlink slots for acknowledgments and scheduling information.
The Wiener processes that the nodes estimate have the same scaling factor $\sigma=4$~Us$^{-1}$, where U is the arbitrary unit measuring the process, e.g., Pascals for a pressure sensor. This assumption will be relaxed in Sec. \ref{sec:res_robustness}, studying the case in which each sensor has a different value of $\sigma_n$. The belief distribution $\varphi_n$ was also discretized to $V=1000$ possible values for computational reasons. The main simulation parameters are listed in Table~\ref{tab:parameters}.

\begin{table}[b]
    \centering
    \caption{Simulation parameters.}
    \begin{tabular}{cc|cc}
        \toprule
        Parameter & Value & Parameter & Value \\
        \midrule
        $N_{\text{run}}$ & $50$~episodes & $L$ & $200$~frames\\
        $F$ & $10$ & $T_s$ & $10$~ms\\
        $S$ & $\{12,22,42,62,82\}$~slots & $N$ & $2000$~nodes\\
        $\sigma$ & $4$~U/s & $V$&$1000$~values\\
        \bottomrule
    \end{tabular}
    \label{tab:parameters}
\end{table}

We compare the performance of \gls{goirsa} against the optimal pull-based solution, which was run with $S=82$, as it was the setting under which it performed best. In the case where $\sigma$ is the same for all nodes, this corresponds to a simple round robin policy, but when we introduce different values of $\sigma_n$ throughout the network, the pull-based schedule ranks nodes by their expected \gls{voi}, following an indexing approach similar to~\cite{ornee2026remote}. We note that, since the pull-based approach is collision-free, its throughput is $1$~packet per \gls{re}, giving it a significant advantage over random access schemes.

A value-unaware \gls{irsa} approach was also considered as a potential benchmark, but its \gls{mse} was at least twice as high as for the pull-based solution in all scenarios, so it is not shown in the plots. As part of our robustness analysis, we will also compare it to a \gls{tirsa} solution with a fixed threshold, similar to the solution in~\cite{mubarak2026on}. We used the replica distribution $\Lambda^{(2)}(x) = 0.86x^3+0.14x^8$ for all \gls{irsa}-based schemes.

\subsection{Throughput Analysis}

\begin{figure}[t]
    \centering
    \begin{tikzpicture}
\begin{axis}[
width=\fwidth,
height=0.32\columnwidth,
xmajorgrids,
ymajorgrids,
grid style={line width=.1pt, draw=gray!10},
tick label style={font=\small},
major grid style={line width=.2pt,draw=gray!40},
xlabel={$G-\bar{G}$},
ylabel={Probability},
xtick style={color=black},
ymin=0, ymax=0.25,
xmin=35, xmax=65,
ytick style={color=black},
legend style={
    draw=white!15!black,
    font=\scriptsize,
    at={(0.01,0.98)},
    anchor=north west,
    /tikz/every even column/.append style={column sep=0.2em}
},
legend columns=1,
ybar,
bar width=1pt,
xtick=data,
xtick={30,35,40,45,50,55,60,65,70},
xticklabels={-0.2,-0.15,-0.1,-0.05,0,0.05,0.1,0.15,0.2},
ytick={0.0,0.05,0.1,0.15,0.2,0.25},
yticklabels={0,0.05,0.1,0.15,0.2,0.25}
]

\addplot[color0,fill=color0!50] table[
    x expr=\coordindex,
    y=y,
    col sep=comma
] {plot/binomial_distribution.csv};

\addplot[color3,fill=color3!50] table[
    x expr=\coordindex,
    y expr=\thisrow{frequency}/1000,
    col sep=comma
] {plot/results/load_estimation_error_histogram.csv};

\legend{IRSA, GO-IRSA};

\end{axis}
\end{tikzpicture}
    \caption{Probability distribution of the difference between the actual and the target load, with $M=400$, $\bar{G}=0.7$, $N=2000$.}
    \label{fig:hist}
\end{figure}
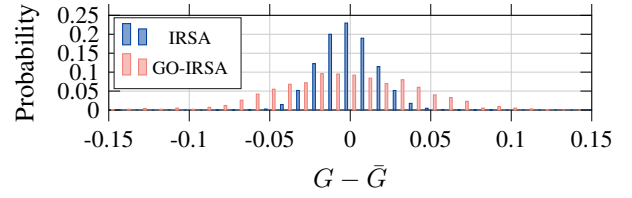

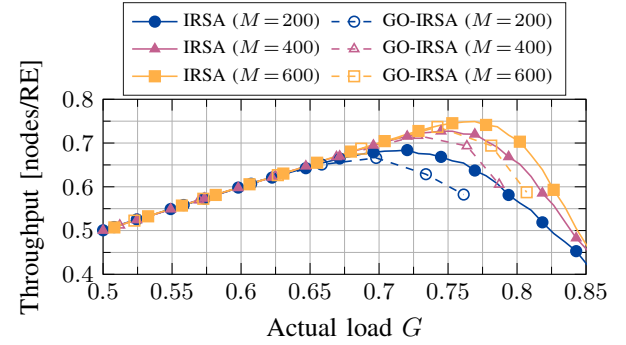
\begin{figure}[t]
    \centering
\begin{tikzpicture}

\definecolor{darkgray176}{RGB}{176,176,176}

\begin{axis}[
width=\fwidth,
height=\fheight,
grid=both,
tick label style={font=\small},
grid style={line width=.1pt, draw=gray!10},
major grid style={line width=.2pt,draw=gray!40},
minor tick num=1,
xlabel={Actual load $G$},
ylabel={Throughput [nodes/RE]},
x grid style={darkgray176},
xmin=0.5, xmax=0.85,
xtick style={color=black},
y grid style={darkgray176},
ymin=0.4, ymax=.8,
ytick style={color=black},
legend style={draw=white!15!black, font=\scriptsize, at={(0.5, 1.02)}, anchor=south, /tikz/every even column/.append style={column sep=0.2em}},
legend columns=2
]

\addplot[color=color0,semithick,mark=*, mark options={solid},mark size=2, mark repeat=3] table[x=load,y=throughput,col sep=comma] {plot/theoretical_irsa/theoretical_throughput_bcsa_200.csv};

\addplot[color=color0,dashed,semithick,mark=o, mark options={solid},mark size=2] table[x=load,y=throughput,col sep=comma] {plot/results/200/bcsa/results_st.csv};

\addplot[color=color2,semithick,mark=triangle*, mark size=2, mark options={solid}, mark repeat=3, mark phase=1] table[x=load,y=throughput,col sep=comma] {plot/theoretical_irsa/theoretical_throughput_bcsa_400.csv};

\addplot[color=color2,semithick,dashed,mark=triangle, mark size=2, mark options={solid}] table[x=load,y=throughput,col sep=comma] {plot/results/400/bcsa/results_mt.csv};

\addplot[color=color4,semithick,mark=square*, mark size=2, mark options={solid}, mark repeat=3, mark phase=2] table[x=load,y=throughput,col sep=comma] {plot/theoretical_irsa/theoretical_throughput_bcsa_600.csv};

\addplot[color=color4,semithick,mark=square,dashed, mark size=2, mark options={solid}] table[x=load,y=throughput,col sep=comma] {plot/results/600/bcsa/results_st.csv};



\legend{IRSA ($M\!=\!200$), GO-IRSA ($M\!=\!200$),  IRSA ($M\!=\!400$), GO-IRSA ($M\!=\!400$), IRSA ($M\!=\!600$), GO-IRSA ($M\!=\!600$)}
\end{axis}
\end{tikzpicture}
    \caption{Throughput achieved by \gls{irsa} and \gls{goirsa}.}
    \label{fig:go_irsa_throughput}
\end{figure}

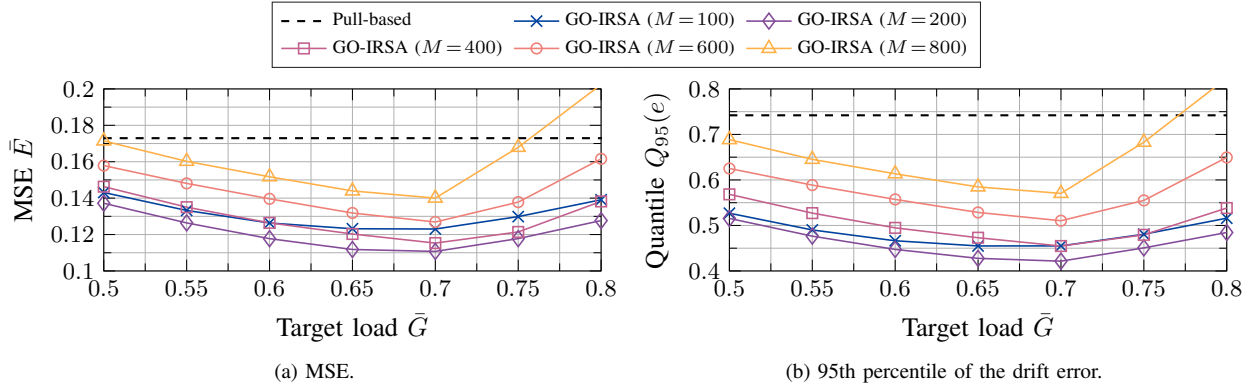
\begin{figure*}
    \centering
    \subfloat{
\begin{tikzpicture}

\definecolor{darkgray176}{RGB}{176,176,176}

\begin{axis}[
width=0cm,
height=0.5cm,
scale only axis,
tick align=inside,
xmin=0, xmax=0,
ymin=0, ymax=0.02,
legend style={
    draw=white!15!black,
    font=\scriptsize, at={(0, 0)}, anchor=south, legend cell align={left},
    /tikz/every even column/.append style={column sep=0.2em}
},
legend columns=3
]

\addplot[black,thick,dashed] table {
0    0.01
};

\addplot[color=color0,semithick,mark=x, mark size=3] table {
0    0.01
};

\addplot[color=color1,semithick,mark=diamond, mark size=3] table {
0    0.01
};
\addplot[color=color2,semithick,mark=square, mark size=2] table {
0    0.01
};

\addplot[color=color3,semithick,mark=o, mark size=2] table {
0    0.01
};

\addplot[color=color4,semithick,mark=triangle, mark size=3] table {
0    0.01
};

\legend{
Pull-based,
GO-IRSA ($M\!=\!100$),
GO-IRSA ($M\!=\!200$),
GO-IRSA ($M\!=\!400$),
GO-IRSA ($M\!=\!600$),
GO-IRSA ($M\!=\!800$)
}

\end{axis}
\end{tikzpicture}} \\ 
    \vspace{-0.3cm}
    \setcounter{subfigure}{0}
    \subfloat[\gls{mse}. \label{fig:go_perf_mse}]{
\begin{tikzpicture}

\definecolor{darkgray176}{RGB}{176,176,176}

\begin{axis}[
width=\twofwidth,
height=\twofheight,
minor tick num=1,
grid=both,
grid style={line width=.1pt, draw=gray!10},
tick label style={font=\small},
major grid style={line width=.2pt,draw=gray!40},
xlabel={Target load $\bar{G}$},
ylabel={MSE $\bar{E}$},
x grid style={darkgray176},
xmin=0.5, xmax=0.8,
xtick style={color=black},
y grid style={darkgray176},
ymin=0.1, ymax=0.2,
ytick style={color=black},
legend style={
    overlay,
    draw=white!15!black,
    font=\scriptsize,
    at={(0.5,1.05)},
    anchor=south,
    /tikz/every even column/.append style={column sep=0.2em}
},
legend columns=2
]

\addplot[color=black,thick,dashed]  coordinates {
    (0, 0.17293925682082772)
    (1, 0.17293925682082772)
};

\addplot[color=color0,semithick,mark=x, mark size=3] table[
    x=target_load,
    y=100,
    col sep=comma
] {plot/results/target_load.csv};

\addplot[color=color1,semithick,mark=diamond, mark size=3] table[
    x=target_load,
    y=200,
    col sep=comma
] {plot/results/target_load.csv};

\addplot[color=color2,semithick,mark=square, mark size=2] table[
    x=target_load,
    y=400,
    col sep=comma
] {plot/results/target_load.csv};

\addplot[color=color3,semithick,mark=o, mark size=2] table[
    x=target_load,
    y=600,
    col sep=comma
] {plot/results/target_load.csv};

\addplot[color=color4,semithick,mark=triangle, mark size=3] table[
    x=target_load,
    y=800,
    col sep=comma
] {plot/results/target_load.csv};

\end{axis}
\end{tikzpicture}}
    \subfloat[95th percentile of the drift error. \label{fig:go_perf_quant}]{
\begin{tikzpicture}

\definecolor{darkgray176}{RGB}{176,176,176}

\begin{axis}[
width=\twofwidth,
height=\twofheight,
grid=both,
grid style={line width=.1pt, draw=gray!10},
major grid style={line width=.2pt,draw=gray!40},
minor tick num=1,
tick label style={font=\small},
xlabel={Target load $\bar{G}$},
ylabel={Quantile $Q_{95}(e)$},
x grid style={darkgray176},
xmin=0.5, xmax=0.8,
xtick style={color=black},
y grid style={darkgray176},
ymin=0.4, ymax=0.8,
ytick style={color=black},
legend style={draw=white!15!black, font=\scriptsize, at={(0.82, 0.98)}, anchor=north east, /tikz/every even column/.append style={column sep=0.2em}},
legend columns=2
]
\addplot[color=black,thick,dashed] coordinates {
    (0, 0.7420408144146203)
    (1, 0.7420408144146203)
};

\addplot[color=color0,semithick,mark=x, mark size=3] table[
    x=target_load,
    y expr=abs(\thisrow{quantile_reward}),
    col sep=comma
] {plot/results/100/bcsa/results_st.csv};

\addplot[color=color1,semithick,mark=diamond, mark size=3] table[
    x=target_load,
    y expr=abs(\thisrow{quantile_reward}),
    col sep=comma
] {plot/results/200/bcsa/results_st.csv};

\addplot[color=color2,semithick,mark=square, mark size=2] table[
    x=target_load,
    y expr=abs(\thisrow{quantile_reward}),
    col sep=comma
] {plot/results/400/bcsa/results_st.csv};

\addplot[color=color3,semithick,mark=o, mark size=2] table[
    x=target_load,
    y expr=abs(\thisrow{quantile_reward}),
    col sep=comma
] {plot/results/600/bcsa/results_st.csv};

\addplot[color=color4,semithick,mark=triangle, mark size=3] table[
    x=target_load,
    y expr=abs(\thisrow{quantile_reward}),
    col sep=comma
] {plot/results/800/bcsa/results_st.csv};

\end{axis}
\end{tikzpicture}}
    \caption{Performance of \gls{goirsa} in the symmetric scenario ($\sigma_n=\sigma\ \forall n$).}\vspace{-0.3cm}
    \label{fig:go_perf}
\end{figure*}

We first consider the throughput that \gls{goirsa} can obtain. Analytical results on \gls{irsa} such as the one in Fig.~\ref{fig:basic_perf_irsa} often assume Poisson arrivals, while we have a limited population of nodes. Assuming a standard \gls{irsa} scheme, the actual load $G$ in a given frame then follows a binomial distribution, which may overshoot the target $\bar{G}$. On the other hand, the load offered by \gls{goirsa} depends on the belief distributions, and may have an even wider variance. Fig.~\ref{fig:hist} shows a simulation of the actual load offered by \gls{irsa} and \gls{goirsa} with the considered $N=2000$ nodes and a target load $\bar{G}=0.7$. We note that, while the load for \gls{irsa} has a small variance, the load offered by \gls{goirsa} has a larger variance, with a long tail and a negative expected value $\E{G-\bar{G}}=-0.013$.

The randomness of the load can affect the performance of the scheme: as Fig.~\ref{fig:plr_irsa} shows, small changes in the load can increase the packet loss rate, especially when operating close to the peak-throughput point, leading to more unresolved frames. This has a negative effect on throughput: Fig.~\ref{fig:go_irsa_throughput} compares the throughput of \gls{irsa} and \gls{goirsa} for different values of $M$, showing that \gls{goirsa} cannot support higher target loads without degrading throughput. As we will discuss in the following, this disadvantage is more than compensated by the goal-oriented nature of \gls{goirsa}, which can ensure that the delivered packets are highly informative and improve the estimation quality even with a slightly reduced throughput. We remark again that operating \gls{irsa} or \gls{goirsa} over longer frames always improves throughput performance due to the more efficient operation of \gls{sic} \cite{Liva11:IRSA,ivanov2017broadcast}.

\subsection{Performance Analysis}

Although throughput represents a major performance indicator for traditional communication systems, the key principle of \gls{goc} is that directly targeting application performance will provide significant benefits.
We thus consider the estimation error of the \gls{dt}, which we measure through the \gls{mse} $\bar{E}$, computed as the average over the whole episode of the frame-level \gls{mse} from~\eqref{eq:mse}, and the worst-case performance $Q_{95}(e)$, corresponding to the $95$th percentile of the error in each frame. Fig.~\ref{fig:go_perf_mse} reports the \gls{mse} of \gls{goirsa} as a function of the target load $\bar{G}$, comparing it to the \gls{mse} obtained by the pull-based scheme.
Considering the optimal working point $\bar{G}^*$ for each configuration, \gls{goirsa} can reduce the \gls{mse} by up to $30\%$, in spite of the collision-free operation of the pull-based approach. We can also note that the schemes are robust to different settings of $\bar{G}$, maintaining a significant gain even if the target load is misconfigured. Similar trends outcome are shown in Fig.~\ref{fig:go_perf_quant} for the worst-case performance.

It is interesting to note that, while targeting throughput would lead to longer frames to reduce the risk of unresolved frames, the performance of \gls{goirsa} is maximized when setting relatively short frames ($M=200$): shorter frames allow nodes to promptly report updates, reducing the risk of high values going unnoticed for a long time and increasing the frequency with which the \gls{dt} gets implicit information about nodes that do not transmit in fully resolved frames. Naturally, there is a trade-off between throughput efficiency and responsiveness, and setting $M=100$ degrades performance due to the high packet loss rate. On the other hand, higher levels of $M$ suffer because of their low responsiveness.



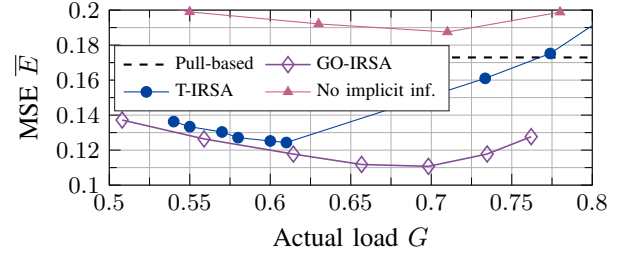
\begin{figure}
    \centering
\begin{tikzpicture}

\definecolor{darkgray176}{RGB}{176,176,176}

\begin{axis}[
width=\fwidth,
height=\fheight,
grid=both,
tick label style={font=\small},
grid style={line width=.1pt, draw=gray!10},
major grid style={line width=.2pt,draw=gray!40},
minor tick num=1,
xlabel={Actual load $G$},
ylabel={MSE $\overline{E}$},
x grid style={darkgray176},
xmin=0.5, xmax=0.8,
xtick style={color=black},
y grid style={darkgray176},
ymin=0.1, ymax=0.2,
ytick style={color=black},
legend style={draw=white!15!black, legend cell align={left}, font=\scriptsize, at={(0.01, 0.8)}, anchor=north west, /tikz/every even column/.append style={column sep=0.2em}},
legend columns=2
]
\addplot[color=black,thick,dashed] coordinates {
    (0, 0.17293925682082772)
    (1.2, 0.17293925682082772)
};

\addplot[color=color1,semithick,mark=diamond, mark size=3] table[
    x=load,
    y expr=abs(\thisrow{average_reward}),
    col sep=comma
] {plot/results/200/bcsa/results_mt.csv};


\addplot[color0,mark=*] table[
    x=load,
    y=error,
    col sep=comma
] {plot/ablation/results_200.csv};

\addplot[color2,mark=triangle*] table[
    x=load,
    y=average_reward,
    col sep=comma
] {plot/ablation/results_800.csv};

\legend{Pull-based, GO-IRSA, T-IRSA , No implicit inf.}
\end{axis}
\end{tikzpicture}
    \caption{Performance of \gls{goirsa} against \gls{tirsa} an adaptive threshold with no implicit information, with $M=200$.}
    \label{fig:fixed_th_comp}
\end{figure}

\begin{figure*}
    \centering
    \subfloat{
\begin{tikzpicture}

\definecolor{darkgray176}{RGB}{176,176,176}

\begin{axis}[
width=0cm,
height=0.1cm,
scale only axis,
tick align=inside,
xmin=0, xmax=0,
ymin=0, ymax=0.1,
legend style={
    draw=white!15!black,
    font=\scriptsize, at={(0, 0)}, anchor=south,
    /tikz/every even column/.append style={column sep=0.2em}
},
legend columns=6
]

\addplot[color=color0, dashdotted,semithick] table {
0    0.01
};

\addplot[color=color1] table {
0    0.01
};

\addplot[color=color2] table {
0    0.01
};

\addplot[color=color3] table {
0    0.01
};

\addplot[color4,thick,densely dashed] table {
0    0.01
};

\legend{
$\bar{G}$,
$G(k)$,
$\theta(k)$,
$E(k)$,
Unresolved frame
}

\end{axis}
\end{tikzpicture}} \\ 
    \vspace{-0.3cm}
    \setcounter{subfigure}{0}    \subfloat[Offered \gls{irsa} load. \label{fig:load_time}]
    {
\begin{tikzpicture}

\definecolor{darkgray176}{RGB}{176,176,176}

\begin{axis}[
width=\threefwidth,
height=\threefheight,
grid=both,
grid style={line width=.1pt, draw=gray!10},
major grid style={line width=.2pt,draw=gray!40},
minor tick num=1,
xlabel={Frame $k$},
tick label style={font=\small},
ylabel={Load $G(k)$},
x grid style={darkgray176},
xmin=0.0, xmax=200,
xtick style={color=black},
y grid style={darkgray176},
ymin=0.4, ymax=1.0,
ytick style={color=black},
legend style={draw=white!15!black, font=\scriptsize, at={(0.7, 0.98)}, anchor=north east, /tikz/every even column/.append style={column sep=0.2em}},
legend columns=3]

\addplot[color4,thick,densely dashed] coordinates {
    (7,0)
    (7,1)
};
\addplot[color4,thick,densely dashed] coordinates {
    (13,0)
    (13,1)
};
\addplot[color4,thick,densely dashed] coordinates {
    (34,0)
    (34,1)
};
\addplot[color4,thick,densely dashed] coordinates {
    (45,0)
    (45,1)
};
\addplot[color4,thick,densely dashed] coordinates {
    (50,0)
    (50,1)
};
\addplot[color4,thick,densely dashed] coordinates {
    (52,0)
    (52,1)
};
\addplot[color4,thick,densely dashed] coordinates {
    (93,0)
    (93,1)
};
\addplot[color4,thick,densely dashed] coordinates {
    (95,0)
    (95,1)
};
\addplot[color4,thick,densely dashed] coordinates {
    (122,0)
    (122,1)
};
\addplot[color4,thick,densely dashed] coordinates {
    (145,0)
    (145,1)
};

\addplot[color1] table[x expr=\coordindex, y=load, col sep=comma] {plot/ablation/good_episode.csv};

\addplot[dashdotted,semithick,color0] coordinates {
(0,0.7)
(200,0.7)
};

\end{axis}
\end{tikzpicture}}
    \subfloat[Transmission threshold. \label{fig:threshold_time}]
    {
\begin{tikzpicture}

\definecolor{darkgray176}{RGB}{176,176,176}

\begin{axis}[
width=\threefwidth,
height=\threefheight,
grid=both,
grid style={line width=.1pt, draw=gray!10},
major grid style={line width=.2pt,draw=gray!40},
minor tick num=1,
xlabel={Frame $k$},
ylabel={Threshold $\theta(k)$},
tick label style={font=\small},
x grid style={darkgray176},
xmin=0.0, xmax=200,
xtick style={color=black},
y grid style={darkgray176},
ymin=40.0, ymax=70.0,
ytick style={color=black},
ytick={40,50,60,70},
yticklabels={0.25,0.5,0.75, 1},
legend style={draw=white!15!black, font=\scriptsize, at={(0.7, 0.98)}, anchor=north east, /tikz/every even column/.append style={column sep=0.2em}},
legend columns=2
]

\addplot[color4,thick,densely dashed] coordinates {
    (7,0)
    (7,100)
};
\addplot[color4,thick,densely dashed] coordinates {
    (13,0)
    (13,100)
};
\addplot[color4,thick,densely dashed] coordinates {
    (34,0)
    (34,100)
};
\addplot[color4,thick,densely dashed] coordinates {
    (45,0)
    (45,100)
};
\addplot[color4,thick,densely dashed] coordinates {
    (50,0)
    (50,100)
};
\addplot[color4,thick,densely dashed] coordinates {
    (52,0)
    (52,100)
};
\addplot[color4,thick,densely dashed] coordinates {
    (93,0)
    (93,100)
};
\addplot[color4,thick,densely dashed] coordinates {
    (95,0)
    (95,100)
};
\addplot[color4,thick,densely dashed] coordinates {
    (122,0)
    (122,100)
};
\addplot[color4,thick,densely dashed] coordinates {
    (145,0)
    (145,100)
};

\addplot[color2] table[x expr=\coordindex, y=threshold, col sep=comma] {plot/ablation/good_episode.csv};

\end{axis}
\end{tikzpicture}}
    \subfloat[Process \gls{mse}. \label{fig:error_time}]
    {
\begin{tikzpicture}

\definecolor{darkgray176}{RGB}{176,176,176}

\begin{axis}[
width=\threefwidth,
height=\threefheight,
grid=both,
grid style={line width=.1pt, draw=gray!10},
major grid style={line width=.2pt,draw=gray!40},
minor tick num=1,
xlabel={Frame $k$},
ylabel={MSE $E(k)$},
tick label style={font=\small},
x grid style={darkgray176},
xmin=0.0, xmax=200,
xtick style={color=black},
y grid style={darkgray176},
ymin=0.05, ymax=0.2,
ytick={0.05,0.1,0.15,0.2},
yticklabels={0.05,0.1,0.15,0.2},
ytick style={color=black},
legend style={draw=white!15!black, font=\scriptsize, at={(0.7, 0.98)}, anchor=north east, /tikz/every even column/.append style={column sep=0.2em}},
legend columns=3]

\addplot[color4,thick,densely dashed] coordinates {
    (7,0)
    (7,1)
};
\addplot[color4,thick,densely dashed] coordinates {
    (13,0)
    (13,1)
};
\addplot[color4,thick,densely dashed] coordinates {
    (34,0)
    (34,1)
};
\addplot[color4,thick,densely dashed] coordinates {
    (45,0)
    (45,1)
};
\addplot[color4,thick,densely dashed] coordinates {
    (50,0)
    (50,1)
};
\addplot[color4,thick,densely dashed] coordinates {
    (52,0)
    (52,1)
};
\addplot[color4,thick,densely dashed] coordinates {
    (93,0)
    (93,1)
};
\addplot[color4,thick,densely dashed] coordinates {
    (95,0)
    (95,1)
};
\addplot[color4,thick,densely dashed] coordinates {
    (122,0)
    (122,1)
};
\addplot[color4,thick,densely dashed] coordinates {
    (145,0)
    (145,1)
};
\addplot[color3] table[x expr=\coordindex, y=error, col sep=comma] {plot/ablation/good_episode.csv};

\end{axis}
\end{tikzpicture}}
    \caption{Performance of \gls{goirsa} during a single episode with target load $\bar{G}=0.7$.}\vspace{-0.4cm}
    \label{fig:go_episode}
\end{figure*}
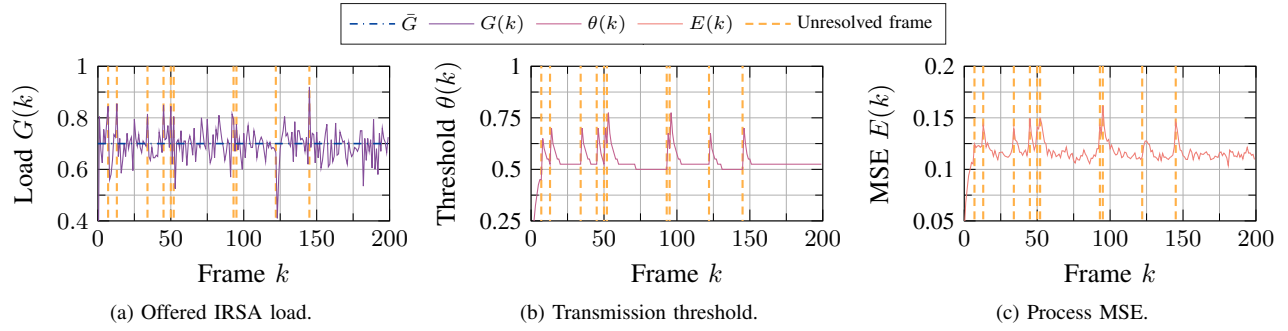

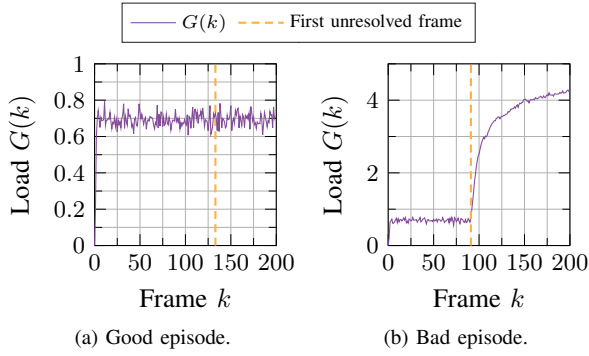
\begin{figure}
    \centering
    \subfloat{
\begin{tikzpicture}

\definecolor{darkgray176}{RGB}{176,176,176}

\begin{axis}[
width=0cm,
height=0.1cm,
scale only axis,
tick align=inside,
xmin=0, xmax=0,
ymin=0, ymax=0.1,
legend style={
    draw=white!15!black,
    font=\scriptsize, at={(0, 0)}, anchor=south,
    /tikz/every even column/.append style={column sep=0.2em}
},
legend columns=6
]

\addplot[color=color1] table {
0    0.01
};

\addplot[color4,thick,densely dashed] table {
0    0.01
};

\legend{
$G(k)$,
First unresolved frame
}

\end{axis}
\end{tikzpicture}} \\ 
    \vspace{-0.3cm}
    \setcounter{subfigure}{0}    
    \subfloat[Good episode. \label{fig:fixed_threshold_time_good}] {
\begin{tikzpicture}

\definecolor{darkgray176}{RGB}{176,176,176}

\begin{axis}[
width=\twofwidth,
height=\twofwidth,
grid=both,
grid style={line width=.1pt, draw=gray!10},
major grid style={line width=.2pt,draw=gray!40},
minor tick num=1,
tick label style={font=\small},
xlabel={Frame $k$},
ylabel={Load $G(k)$},
x grid style={darkgray176},
xmin=0.0, xmax=200,
xtick style={color=black},
y grid style={darkgray176},
ymin=0, ymax=1.0,
ytick style={color=black},
legend style={draw=white!15!black, font=\scriptsize, at={(0.7, 0.98)}, anchor=north east, /tikz/every even column/.append style={column sep=0.2em}},
legend columns=3]

\addplot[color4,thick,densely dashed] coordinates {
    (133,0)
    (133,4.25)
};

\addplot[color1] table[x expr=\coordindex, y=load, col sep=comma] {plot/ablation/good_episode_fixed.csv};

\end{axis}
\end{tikzpicture}}
    \subfloat[Bad episode. \label{fig:fixed_threshold_time_bad}] {
\begin{tikzpicture}

\definecolor{darkgray176}{RGB}{176,176,176}

\begin{axis}[
width=\twofwidth,
height=\twofwidth,
tick label style={font=\small},
grid=both,
grid style={line width=.1pt, draw=gray!10},
major grid style={line width=.2pt,draw=gray!40},
minor tick num=1,
xlabel={Frame $k$},
ylabel={Load $G(k)$},
x grid style={darkgray176},
xmin=0.0, xmax=200,
xtick style={color=black},
y grid style={darkgray176},
ymin=0, ymax=5,
ytick style={color=black},
legend style={draw=white!15!black, font=\scriptsize, at={(0.7, 0.98)}, anchor=north east, /tikz/every even column/.append style={column sep=0.2em}},
legend columns=3]

\addplot[color1] table[x expr=\coordindex, y=load, col sep=comma] {plot/ablation/bad_episode.csv};

\addplot[color4,thick,densely dashed] coordinates {
    (91,0)
    (91,5)
};

\end{axis}
\end{tikzpicture}}
    \caption{Actual load $G$ during a single episode with fixed threshold.}
    \label{fig:fix_episode}
\end{figure}

We then compare the proposed \gls{goirsa} scheme against two other \gls{irsa}-based goal-oriented schemes: the first is the \gls{tirsa} scheme using a fixed threshold inspired by~\cite{mubarak2026on}, while the second is a modified version of \gls{goirsa} that does not use implicit information. In this scheme, the posterior belief distribution is only updated for nodes who successfully transmit, while all other nodes are updated as in unresolved frames. In other words, the ``silence'' branch in the diagram from Fig.~\ref{fig:belief_update} is removed, and all nodes that did not have a successful transmission go through the ``unresolved'' branch. This comparison is shown in Fig.~\ref{fig:fixed_th_comp}. We note that, as the concept of target load cannot be applied to \gls{tirsa}, we plotted the results against the actual load $G$, i.e., the average number of sensors transmitting per RE as measured in the simulations. We also set $M=200$ for all \gls{irsa}-based policies: in the \gls{tirsa} case, this corresponds to the optimal frame length, while the version with no implicit information would benefit from longer frames, although its performance is always worse than the pull-based scheme. This is an indication of the importance of implicit information: without these updates, \gls{goirsa} would not have a performance advantage, as its belief distribution $\varphi_n$ would be overly pessimistic and misleading. On the other hand, \gls{tirsa} performs rather well when considering lower values of the load $G$, but there is a critical point over which decreasing the threshold by even a small amount leads to significant performance degradation and increased load: setting $\theta=0.6585$, we get a load $G\sim0.6$, while setting $\theta=0.6582$, the load is over $0.8$, with a significant increase in the \gls{mse}. As we will discuss below, this is caused by its fundamental instability, which appears as soon as the packet loss rate becomes non-negligible.

We can better understand the advantages of \gls{goirsa} by examining Fig.~\ref{fig:go_episode}, which illustrates the performance of the proposed approach over a single episode. As shown by Fig.~\ref{fig:load_time}, \gls{goirsa} succeeds in maintaining the system load $G$ close to the target value $\bar{G}=0.7$: after an unresolved frame, i.e., a frame in which the \gls{bs} is not able to successfully decode all packets, $G$ tends to drop significantly, quickly returning to the target value after a couple of frames. Fig.~\ref{fig:threshold_time} provides an explanation for this: \gls{goirsa} immediately increases the threshold $\theta$ after each unresolved frame, as not being able to use implicit information increases its uncertainty. This reduced load allows collided nodes, which presumably have high-value information, to successfully send their updates, and the system quickly returns to the baseline. The \gls{mse} performance, shown in Fig.~\ref{fig:error_time}, reflects the threshold changes, with some spikes corresponding to the unresolved frames, from which the performance quickly recovers, remaining stable for long periods in which all frames are fully resolved.

As this recovery is enabled by dynamically adapting the threshold, it is obviously not possible for \gls{tirsa}: under a fixed threshold, any unresolved frame leads to increased load in the next frame, as more nodes get higher values and attempt to transmit, which, in turn, increases the probability that the next frame will be unresolved as well. This positive feedback loop causes a snowball effect that is clear from Fig.~\ref{fig:fix_episode}, reporting the load generated by \gls{tirsa} in two episodes. In the first one, shown in Fig.~\ref{fig:fixed_threshold_time_good}, the evolution of the Wiener process allows for recovery, but in the second, shown in Fig.~\ref{fig:fixed_threshold_time_bad}, the snowball effect quickly leads to an unsustainable load, as more and more nodes try to transmit. This effect is the cause of the instability we discussed above, and makes its performance highly sensitive to system parameters.

\subsection{Robustness Analysis}
\label{sec:res_robustness}

\begin{figure}
    \centering
\begin{tikzpicture}

\definecolor{darkgray176}{RGB}{176,176,176}

\begin{axis}[
width=\fwidth,
height=\fheight,
tick label style={font=\small},
grid=both,
grid style={line width=.1pt, draw=gray!10},
major grid style={line width=.2pt,draw=gray!40},
minor tick num=1,
xlabel={Target load $\bar{G}$},
ylabel={MSE $\overline{E}$},
x grid style={darkgray176},
xmin=0.5, xmax=0.8,
xtick style={color=black},
y grid style={darkgray176},
ymin=0.1, ymax=0.2,
ytick style={color=black},
legend style={
    draw=white!15!black,
    font=\scriptsize,
    at={(0.5,1.05)},
    legend cell align={left},
    anchor=south,
    /tikz/every even column/.append style={column sep=0.2em}
},
legend columns=2
]

\addplot[black,thick,dashed] coordinates {
    (0, 0.17293925682082772)
    (1, 0.17293925682082772)
};

\addplot[color=color0,semithick,mark=x, mark size=3] table[
    x=target_load,
    y=average_reward,
    col sep=comma
] {plot/imperfect_sic/imperfect_sic_100_slots.csv};

\addplot[color=color1,semithick,mark=diamond, mark size=3] table[
    x=target_load,
    y=average_reward,
    col sep=comma
] {plot/imperfect_sic/imperfect_sic_200_slots.csv};

\addplot[color=color2,semithick,mark=square, mark size=2] table[
    x=target_load,
    y=average_reward,
    col sep=comma
] {plot/imperfect_sic/imperfect_sic_400_slots.csv};

\addplot[color=color3,semithick,mark=o, mark size=2] table[
    x=target_load,
    y=average_reward,
    col sep=comma
] {plot/imperfect_sic/imperfect_sic_600_slots.csv};

\addplot[color=color4,semithick,mark=triangle, mark size=3] table{
0.5 0.2132
0.55 0.2046
0.6 0.1974
0.65 0.1912
0.7 0.1882
0.75 0.1876
0.8 0.2196
};

\legend{
Pull-based,
GO-IRSA ($M\!=\!100$),
GO-IRSA ($M\!=\!200$),
GO-IRSA ($M\!=\!400$),
GO-IRSA ($M\!=\!600$),
GO-IRSA ($M\!=\!800$)
}

\end{axis}
\end{tikzpicture}
    \caption{Performance of \gls{goirsa} with imperfect \gls{sic} ($\eta=0.95$).}
    \label{fig:imperfect_sic}
\end{figure}
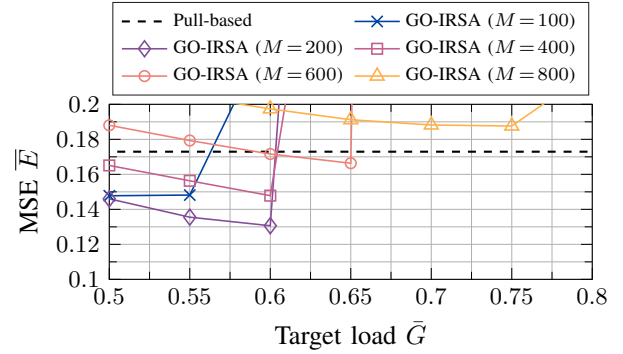

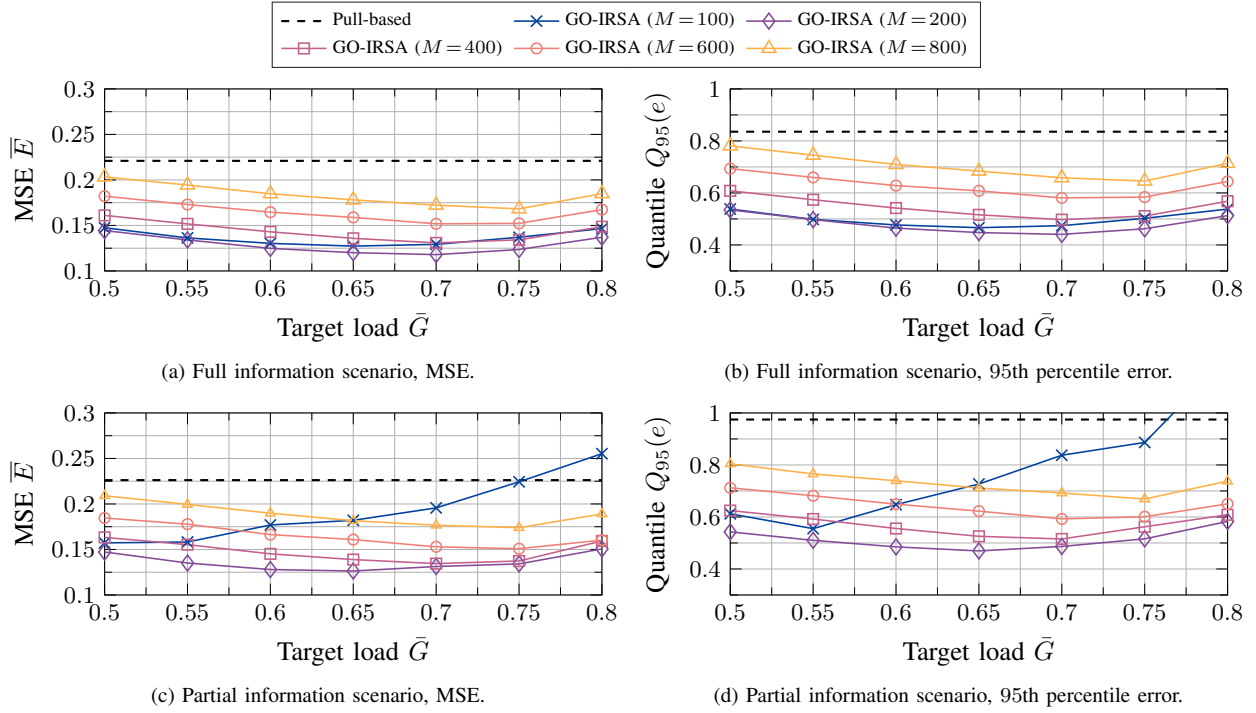
\begin{figure*}
    \centering
    \subfloat{
\begin{tikzpicture}

\definecolor{darkgray176}{RGB}{176,176,176}

\begin{axis}[
width=0cm,
height=0.5cm,
scale only axis,
tick align=inside,
xmin=0, xmax=0,
ymin=0, ymax=0.02,
legend style={
    draw=white!15!black,
    font=\scriptsize, at={(0, 0)}, anchor=south, legend cell align={left},
    /tikz/every even column/.append style={column sep=0.2em}
},
legend columns=3
]

\addplot[black,thick,dashed] table {
0    0.01
};

\addplot[color=color0,semithick,mark=x, mark size=3] table {
0    0.01
};

\addplot[color=color1,semithick,mark=diamond, mark size=3] table {
0    0.01
};
\addplot[color=color2,semithick,mark=square, mark size=2] table {
0    0.01
};

\addplot[color=color3,semithick,mark=o, mark size=2] table {
0    0.01
};

\addplot[color=color4,semithick,mark=triangle, mark size=3] table {
0    0.01
};

\legend{
Pull-based,
GO-IRSA ($M\!=\!100$),
GO-IRSA ($M\!=\!200$),
GO-IRSA ($M\!=\!400$),
GO-IRSA ($M\!=\!600$),
GO-IRSA ($M\!=\!800$)
}

\end{axis}
\end{tikzpicture}} \\ 
    \vspace{-0.3cm}
    \setcounter{subfigure}{0}
    \subfloat[Full information scenario, \gls{mse}.]{
\begin{tikzpicture}

\definecolor{darkgray176}{RGB}{176,176,176}

\begin{axis}[
width=\twofwidth,
height=\twofheight,
tick label style={font=\small},
grid=both,
grid style={line width=.1pt, draw=gray!10},
major grid style={line width=.2pt,draw=gray!40},
minor tick num=1,
xlabel={Target load $\bar{G}$},
ylabel={MSE $\overline{E}$},
x grid style={darkgray176},
xmin=0.5, xmax=0.8,
xtick style={color=black},
y grid style={darkgray176},
ymin=0.1, ymax=0.3,
ytick style={color=black},
legend style={
    draw=white!15!black,
    font=\scriptsize,
    at={(0.95,0.2)},
    anchor=north east,
    /tikz/every even column/.append style={column sep=0.2em}
},
legend columns=2
]

\addplot[black,thick,dashed] coordinates {
    (0, 0.22095818850398063)
    (1, 0.22095818850398063)
};

\addplot[color=color0,semithick,mark=x, mark size=3] table[
    x=target_load,
    y=average_reward,
    col sep=comma
] {plot/different_sigmas/results_100.csv};

\addplot[color=color1,semithick,mark=diamond, mark size=3] table[
    x=target_load,
    y=average_reward,
    col sep=comma
] {plot/different_sigmas/results_200.csv};

\addplot[color=color2,semithick,mark=square, mark size=2] table[
    x=target_load,
    y=average_reward,
    col sep=comma
] {plot/different_sigmas/results_400.csv};

\addplot[color=color3,semithick,mark=o, mark size=2] table[
    x=target_load,
    y=average_reward,
    col sep=comma
] {plot/different_sigmas/results_600.csv};

\addplot[color=color4,semithick,mark=triangle, mark size=3] table{
0.5 0.2033
0.55 0.1944
0.6 0.1848
0.65 0.1781
0.7 0.1723
0.75 0.1682
0.8 0.1850
};

\end{axis}
\end{tikzpicture}\label{fig:go_irsa_different_sigma_mse}}
    \subfloat[Full information scenario, $95$th percentile error.]{
\begin{tikzpicture}

\definecolor{darkgray176}{RGB}{176,176,176}

\begin{axis}[
width=\twofwidth,
height=\twofheight,
tick label style={font=\small},
grid=both,
grid style={line width=.1pt, draw=gray!10},
major grid style={line width=.2pt,draw=gray!40},
minor tick num=1,
xlabel={Target load $\bar{G}$},
ylabel={Quantile $Q_{95}(e)$},
x grid style={darkgray176},
xmin=0.5, xmax=0.8,
xtick style={color=black},
y grid style={darkgray176},
ymin=0.3, ymax=1.0,
ytick style={color=black},
legend style={draw=white!15!black, font=\scriptsize, at={(0.82, 0.98)}, anchor=north east, /tikz/every even column/.append style={column sep=0.2em}},
legend columns=2
]
\addplot[color=black,thick,dashed] coordinates {
    (0, 0.835492704808712)
    (1, 0.835492704808712)
};

\addplot[color=color0,semithick,mark=x, mark size=3] table[
    x=target_load,
    y expr=abs(\thisrow{quantile_reward}),
    col sep=comma
] {plot/different_sigmas/results_100.csv};

\addplot[color=color1,semithick,mark=diamond, mark size=3] table[
    x=target_load,
    y expr=abs(\thisrow{quantile_reward}),
    col sep=comma
] {plot/different_sigmas/results_200.csv};

\addplot[color=color2,semithick,mark=square, mark size=2] table[
    x=target_load,
    y expr=abs(\thisrow{quantile_reward}),
    col sep=comma
] {plot/different_sigmas/results_400.csv};

\addplot[color=color3,semithick,mark=o, mark size=2] table[
    x=target_load,
    y expr=abs(\thisrow{quantile_reward}),
    col sep=comma
] {plot/different_sigmas/results_600.csv};

\addplot[color=color4,semithick,mark=triangle, mark size=3] table{
0.5 0.7807914496958256
0.55 0.7452367818355561
0.6 0.7093496388196945
0.65 0.6839938935637474
0.7 0.6583441197872162
0.75 0.6457216829061508
0.8 0.7138112270832062
};

\end{axis}
\end{tikzpicture}\label{fig:go_irsa_different_sigma_quant}}\\ \vspace{-0.2cm}
    \subfloat[Partial information scenario, \gls{mse}. ]{
\begin{tikzpicture}

\definecolor{darkgray176}{RGB}{176,176,176}

\begin{axis}[
width=\twofwidth,
height=\twofheight,
tick label style={font=\small},
grid=both,
grid style={line width=.1pt, draw=gray!10},
major grid style={line width=.2pt,draw=gray!40},
minor tick num=1,
xlabel={Target load $\bar{G}$},
ylabel={MSE $\overline{E}$},
x grid style={darkgray176},
xmin=0.5, xmax=0.8,
xtick style={color=black},
y grid style={darkgray176},
ymin=0.1, ymax=0.3,
ytick style={color=black},
legend style={draw=white!15!black, font=\scriptsize, at={(0.82, 0.98)}, anchor=north east, /tikz/every even column/.append style={column sep=0.2em}},
legend columns=2
]
\addplot[color=black,thick,dashed] coordinates {
    (0, 0.22618581584095956)
    (1, 0.22618581584095956)
};

\addplot[color=color0,semithick,mark=x, mark size=3] table[
    x=target_load,
    y expr=abs(\thisrow{average_reward}),
    col sep=comma
] {plot/different_sigmas/no_info_100.csv};

\addplot[color=color1,semithick,mark=diamond, mark size=3] table[
    x=target_load,
    y expr=abs(\thisrow{average_reward}),
    col sep=comma
] {plot/different_sigmas/no_info_200.csv};

\addplot[color=color2,semithick,mark=square, mark size=2] table[
    x=target_load,
    y expr=abs(\thisrow{average_reward}),
    col sep=comma
] {plot/different_sigmas/no_info_400.csv};

\addplot[color=color3,semithick,mark=o, mark size=2] table[
    x=target_load,
    y expr=abs(\thisrow{average_reward}),
    col sep=comma
] {plot/different_sigmas/no_info_600.csv};

\addplot[color=color4,semithick,mark=triangle, mark size=2] table[
    x=target_load,
    y expr=abs(\thisrow{average_reward}),
    col sep=comma
] {plot/different_sigmas/no_info_800.csv};

\end{axis}
\end{tikzpicture}\label{fig:no_info_different_sigma_mse}}
    \subfloat[Partial information scenario, $95$th percentile error. ]{
\begin{tikzpicture}

\definecolor{darkgray176}{RGB}{176,176,176}

\begin{axis}[
width=\twofwidth,
height=\twofheight,
tick label style={font=\small},
grid=both,
grid style={line width=.1pt, draw=gray!10},
major grid style={line width=.2pt,draw=gray!40},
minor tick num=1,
xlabel={Target load $\bar{G}$},
ylabel={Quantile $Q_{95}(e)$},
x grid style={darkgray176},
xmin=0.5, xmax=0.8,
xtick style={color=black},
y grid style={darkgray176},
ymin=0.3, ymax=1.0,
ytick style={color=black},
legend style={draw=white!15!black, font=\scriptsize, at={(0.82, 0.98)}, anchor=north east, /tikz/every even column/.append style={column sep=0.2em}},
legend columns=2
]
\addplot[color=black,thick,dashed] coordinates {
    (0, 0.9746874787211418)
    (1, 0.9746874787211418)
};

\addplot[color=color0,semithick,mark=x, mark size=3] table[
    x=target_load,
    y expr=abs(\thisrow{quantile_reward}),
    col sep=comma
] {plot/different_sigmas/no_info_100.csv};

\addplot[color=color1,semithick,mark=diamond, mark size=3] table[
    x=target_load,
    y expr=abs(\thisrow{quantile_reward}),
    col sep=comma
] {plot/different_sigmas/no_info_200.csv};

\addplot[color=color2,semithick,mark=square, mark size=2] table[
    x=target_load,
    y expr=abs(\thisrow{quantile_reward}),
    col sep=comma
] {plot/different_sigmas/no_info_400.csv};

\addplot[color=color3,semithick,mark=o, mark size=2] table[
    x=target_load,
    y expr=abs(\thisrow{quantile_reward}),
    col sep=comma
] {plot/different_sigmas/no_info_600.csv};

\addplot[color=color4,semithick,mark=triangle, mark size=2] table[
    x=target_load,
    y expr=abs(\thisrow{quantile_reward}),
    col sep=comma
] {plot/different_sigmas/no_info_800.csv};

\end{axis}
\end{tikzpicture}\label{fig:no_info_percentile_different_sigma_quant}}
    \caption{Performance of \gls{goirsa} with variable $\sigma_n\in\mc{U}\left(\frac{\sigma}{2},\frac{3\sigma}{2}\right)$.}\label{fig:go_irsa_different_sigma}
\end{figure*}

In the following, we evaluate the robustness of \gls{goirsa} when some of the assumptions in the above are violated. We first consider the possibility that the \gls{sic} process does not guarantee perfect decoding due to the wireless channel. Specifically, after the waveform for each decoded packet is subtracted from slots with other replicas, we consider the possibility that the received signal may still be too degraded to retrieve additional packets. Each round of \gls{sic} then succeeds with probability $\eta=0.95$. This model is often used to easily represent realistic \gls{sic} in \gls{irsa} \cite{dumas2021design} and in other non-orthogonal communication schemes.
Fig.~\ref{fig:imperfect_sic} reports the performance of \gls{goirsa} in this scenario. The large number of unresolved frames due to imperfect \gls{sic} reduces the robustness of the scheme, limiting its applicability to lower target loads. However, we still obtain a $25\%$ gain over the pull-based scheme when setting $\bar{G}^*=0.6$.

Finally, we consider an asymmetric case in which nodes do not have the same $\sigma$. We draw the scaling parameter for each node from a uniform distribution in $[\sigma/2,3\sigma/2]$, i.e., $\sigma_n\sim\mc{U}(\sigma/2,3\sigma/2)$, and we consider two cases. In the \emph{full information} scenario, the exact value of $\sigma_n$ for each node is known to the \gls{bs}, while in the \emph{partial information} scenario, the BS only knows $\sigma$, and thus has an imperfect knowledge of the statistics of the Wiener processes.

The results for this scenario are shown in Fig.~\ref{fig:go_irsa_different_sigma}. We first consider performance in the full information scenario, shown in Fig.~\ref{fig:go_irsa_different_sigma_mse}-\subref*{fig:go_irsa_different_sigma_quant}: even when using the optimal indexing policy, the pull-based benchmark's performance degrades,
as the \gls{mse} increases from $\overline{E}\simeq0.17$ to $\overline{E}\simeq0.22$, while the worst-case error increases from $Q_{\text{95}}(e)\simeq0.74$ in the symmetric case to $Q_{\text{95}}(e)\simeq0.84$. On the other hand, \gls{goirsa}'s performance is similar to the symmetric case: if we consider the system with $M=200$ and $\bar{G}^*=0.7$ as a target load, the \gls{mse} is $\bar{E}\simeq0.12$ in both cases, while the error percentile goes from $Q_{\text{95}}(e)=0.42$ in the symmetric case to $Q_{\text{95}}(e)=0.44$. The robustness to misconfigurations of $\bar{G}$ and $M$ is also preserved, as the scheme is insensitive to small errors in its settings.

We can note a similar trend even in the partial information case, shown in Fig.~\ref{fig:no_info_different_sigma_mse}-\subref*{fig:no_info_percentile_different_sigma_quant}: in this case, the pull-based scheme corresponds to a round robin policy, increasing its \gls{mse} from $\bar{E}\simeq0.221$ in the full information case to $\bar{E}\simeq0.226$ and its worst-case error from $Q_{\text{95}}(e)\simeq0.84$ to  $Q_{\text{95}}(e)\simeq0.97$. On the other hand, \gls{goirsa} is still extremely robust, as the best \gls{mse} goes from $\bar{E}\simeq0.118$ in the full information case to $\bar{E}\simeq0.126$, while the worst-case error goes from $Q_{\text{95}}(e)\simeq0.44$ to  $Q_{\text{95}}(e)\simeq0.47$. However, we note an interesting effect: the optimal target load $\bar{G}^*=0.7$ for the full information case, but $\bar{G}^*=0.65$ under partial information, and performance degrades for higher target loads. This is particularly evident for $M=100$, whose performance is strongly degraded, and which is even worse than the pull-based schemes if $\bar{G}>0.75$. As the \gls{dt} has an imperfect model of the Wiener process, the belief distributions for nodes become less accurate, and while the threshold is generally correct for lower loads, this inaccuracy can lead to more unresolved frames for higher target loads. As shorter frames are more vulnerable to variations in the target loads and have higher packet loss rates, this phenomenon is particularly evident for $M=100$, but it is also noticeable for $M=200$. However, the \gls{goirsa} scheme still preserves its significant performance gain over pull-based solutions, as well as its robustness to small changes in the parameter settings.

\section{Conclusions}
\label{sec:conclusions}

In this work, we introduced \gls{goirsa}, a new \gls{goc} protocol for push-based massive access. \gls{goirsa} operates over realistic frame structures, without requiring either extremely frequent feedback or complex calculations on the sensor side. 
This framework integrates \gls{irsa} with a goal-oriented approach, which allows the \gls{bs} to maintain a belief over the potential \gls{voi} for each node, setting and broadcasting a threshold for nodes to transmit and exploiting implicit information from silent nodes.
Our simulation results show that \gls{goirsa} consistently outperform conventional pull-based strategies, reducing the average and worst-case error by about $30\%$ and providing robustness to various system imperfections. 

To the best of our knowledge, this work represents one of the first attempts to extend push-based \gls{goma} protocols to massive access scenarios.
In future work, we will focus on validating the proposed framework in more realistic communication environments and protocols such as \gls{nbiot}.

\section*{Use of AI Disclosure}

No AI system was used in any aspect of the present work that affects the scientific conclusions of the paper.

\bibliographystyle{IEEEtran}
\bibliography{bibliography}

\end{document}